THEORETICAL STUDY OF THE SOFT OPTIC MODE SCATTERING IN A RELAXOR FERROELECTRIC: THE STRONG EFFECT OF THE DEPOLARIZING FIELDS

E.Iolin[1]

Physics Department, Lehigh University, Bethlehem, PA 18015 (USA)

eui206@lehigh.edu

ABSTRACT

We analyze scattering of the transverse optic modes by spherically symmetric Polarized Nano Regions (PNR) in the paraelectric phase of relaxor ferroelectrics. Calculations have done in the frame of mean-field model early supposed by E.Iolin & J.Toulouse but depolarization field effects, DF, were taken into account. Elementary excitations of the system are found to be of two types – Vortex (V) and Quasi Polar (QP). DF decreases temperature of the local QP phase transition. The local phase transition temperature is found to be lower (higher) for Vortex than for QP excitations for the case of small (large) size PNR. Therefore both QP and Vortex condensations are possible. Depolarization field hasn't effect on the Vortex scattering. For the case of QP scattering DF effect is essential especially in the case when total angular momentum *j*=1. Vortex and QP (at j ≥2) cross section of scattering Σ are nullified at the small value of momentum q of the incident wave. However QP cross section Σ(j=1) is saturated at q~0.01-0.001, and nullified only at the ultra small value of q corresponding macroscopic value of the soft mode wavelength. This scattering leads to the suppression of the TO spreading wave regime and allows qualitative understand some features of so-called waterfall phenomena observed early by the Brookhaven group in the neutron inelastic scattering in relaxors.

## 1. INTRODUCTION

Over the past several decades, serious progress has been made in understanding the physical characteristics of the disordered ferroelectrics so-called relaxors (see for example reviews [1], [2]). These technically important materials (PMN, PZN, KTN and other) are characterized by chemical and structural disordering. Early on, Burns and Dacol [3] introduced the concept of Polarized Nano Regions (PNR) – small ordered (polarized) regions embedded into the host crystal - as the main feature of relaxors. Since then, the validity of this concept has been confirmed by numerous optical, diffuse and inelastic neutron scattering studies (see for example [4, 5, 6]).
An important observation was made by Gehring et al [6, 7] in their study of the Transverse Optic mode (soft-mode) dispersion in the relaxor PZN-8%PT in a temperature range where PNR's should certainly exist. These authors found that, while perfectly well defined TO peaks were observed in the neutron scattering spectra (constant Q-scan) for momenta $q>q_0$, these peaks disappeared for $q<q_0$ and the value of $1/q_0$ was comparable to the expected PNR size of a few tens of angstroms. In addition, the dispersion curves had the shape of a "waterfall". Different explanations of this phenomenon have been proposed

[1] E-mail: eui206@lehigh.edu or eiolin@netzero.net

[7 ÷ 9]. Gehring et al [9] attributed their results to a sharp step-like increase in the TO damping for $q<q_0$. However, no firm consensus has been reached as to the definite origin of the waterfall phenomenon [8]. Another motivation for the present theoretical work are neutron scattering measurement results of the transverse acoustic mode (TA) dispersion and damping in the relaxor $K_{1-x}Ta_xNbO_3$ (KTN) [10, 11, 12]. The shape of the TA peaks exhibit clear deviations from a standard Lorentzian form (Q-scans) at several TA momentum. These deviations appear to be more or less regular, not easily explained by experimental uncertainty and could be interpreted as evidence for resonant scattering of the TA by PNRs.

It is now generally accepted that polarization correlations (PNR) appear at the so-called Burns temperature, $T_B$, upon cooling. It has also been argued [13] that the correlations appearing at $T=T_B$ are dynamic and become static at a significantly lower temperature, $T^*$, corresponding to a local distortion or local phase transition. $T^*$ is in fact the temperature at which elastic diffuse scattering appears in neutron spectra as well as the characteristic relaxor behavior or frequency dispersion of the dielectric constant. Consequently, the term *polar nanodomains* (PND) has been proposed to describe these polarization regions. So far, we do not have detail information about the shape of these regions, the nature of their boundaries (sharp or smooth), the value of the surface energy etc.

We study PNR effect at the soft mode dynamic in the temperature interval $T^*<T<T_B$ in preposition that the relaxor can be approximated as a two-component system. Our model is the same as we studied before ([14], [15]) - a spherically symmetric PNR embedded in an isotropic medium. However our new analysis is including previously omitted effect of the depolarizing fields (DF) at the transverse optic phonon (TO) scattering by PNR. We found that DF effect is very strong in our case and leads to the qualitative understanding of some early misunderstood feathers of waterfall phenomenon. Gehring et al [9] attributed waterfall results to a sharp step-like increase in the TO damping for the momentum $q<q_0$ .

TOW is corresponded to the angular momentum $\boldsymbol{j\geq 1}$ for the case of isotropic medium. It is known that the cross section of the slow particles scattering, σ, $\boldsymbol{\sigma \sim q^{4j}}$ [17]. Therefore cross section, $\boldsymbol{\sigma}$, of the TO scattering by PNR should be small at the small momentum $\boldsymbol{q}$ and waterfall should be suppressed in this case. Calculation results [14, 15] demonstrate that TO scattering by PNR is really strongly increased at small $\boldsymbol{q}$ and nullified at, say, $\boldsymbol{q}$<0.01 r.l.u. in agreement with general theory [17]. We found that depolarization field effect also leads to the result $\boldsymbol{\sigma \sim q^{4j}}$ but only at the extremely small momentum, $\boldsymbol{q<5E\text{-}7}$ r.l.u., corresponding to the size 1/q ~100 μm. It is happened due to the following. Depolarization field interact with optical mode and created corresponding displacements, which could be titled as Depolarization mode (DM). DM space dependence (~$\boldsymbol{r^{j-1}}$ inside PNR and ~$\boldsymbol{r^{-j(j+1)-1}}$ outside of PNR) is quite different from ~sin($\boldsymbol{qr-\omega t}$) corresponding TO wave. The incident TO wave excites charge fluctuations at the PNR surface. These charge fluctuations create DF interacted with optical mode displacements and leading to the increasing of scattering. We found that scattering cross section $\boldsymbol{\sigma}$ corresponding angular harmonic $\boldsymbol{j}$=1, $\boldsymbol{\sigma_{QP}}$(j=1), isn't nullified at very small momentum $\boldsymbol{q}$ , say $\boldsymbol{q}$ ~ 0.001. Value of $\boldsymbol{\sigma_{QP}}(j=1)$ is steadily increased parallel to the decreasing value of momentum $\boldsymbol{q}$, saturated at the high level (plateau) and further nullified at the extremely small value of momentum, $q$~1E-6 ÷ 1E-7 after passing of the extremely narrow resonance peak of scattering at $\boldsymbol{q=q_{res}}$. Scattering cross section is crossing to the dependence $\boldsymbol{\sigma \sim q^4}$ at $\boldsymbol{q< q_{res}}$. For example value of $\boldsymbol{q_{res}}$=5E-7 r.l.u. and resonance energy width ~1E-12 meV~1Hz. Probably observation of similar resonances is out of limit today experimental technique, taking into account problems with resolution, sample quality and so on. Scattering at $\boldsymbol{q}$~0.01÷0.1 r.l.u. is very far from resonance and so no sensitive for the quality of sample.

Therefore waterfall isn't demised at the practically important very small momentum $\boldsymbol{q}$ if DF isn't suppressed due to the presence of conductivity and so on reasons.

In general problem of the DF is old and considered in any standard book concerning electric features of dielectric or ferroelectric materials [18]. However the interest to the DF effect is strongly increased during last decades due to the creation and research of materials contained many internal boundaries.

The effect of DF is dependent vs. conductivity, screening and so on parameters. An interesting analysis of these numerous questions and necessary references could be found in the paper [19]. DF is directed against the direction of the electric dipoles in the simple cases [18] and therefore leads to the increasing of the free energy. Recently it was shown [20 – 22] that DF contribution to the free energy can lead to the complicated structure of the ground state, instead of simple ferroelectric we could expect for example the presence of the toroidal momentum of polarization.

However the complicated geometry of the polarization displacement dynamically appears even at the temperature which is above temperature of the local phase transition. We found that elementary transverse excitations of the system can be divided into two categories – Vortex (V) and quasi polar (QP[2]). Vortex excitations correspond to closed or loop polarization lines of TO displacements and QP contain open TO polarization lines. Depolarization field has no effect at the Vortex scattering.

In the following parts of paper we formulate the model, similar to previously proposed [14, 15], and provide its new solution for the boundary conditions taking into depolarizing modes.

Our model - a spherically symmetric PNR embedded in an isotropic medium - is the simplest.

We assume that all Hamiltonian (Lagrangian) parameters are the same outside (index 1) and inside (index 2) the PNR, with exception of the temperature of the phase transition (Tc1, Tc2, Tc1< Tc2) and the energy gaps $\omega_0(1)$, $\omega_0(2)$ of the TOW soft-mode at zero momentum, q=0. We also assume that all parameters are constant. Mixing of transverse and longitudinal waves at the PNR surface is taken into account but we assume the same gap parameter for the LO and TO in order to avoid using higher derivatives in our equations. In the Lagrangian, we therefore include the interaction of the polarization with the electromagnetic field and take into account longitudinal Coulomb forces. However, we describe the system with a critical TO and non critical LO behavior because the LO does not usually soften in ferroelectrics and because Coulomb forces tend to raise its frequency at small momentum in ionic crystals, leading to a large energy gap $\Omega_p >> \omega_0(1)$, $\omega_0(2)$ and only a moderate (non-critical) dependence on temperature. LO waves are virtually excited ($\Omega_p >> \omega_0(1)$, $\omega_0(2)$ ) but only in a thin layer near the PNR surface. Depolarizing field has no effect at the Vortex dynamic but is essential for the case of QP excitations. We show that a local instability ("phase transition in PNR") occurs at a temperature $T_L<Tc_2$ where the mode frequency goes to zero locally, $\omega_L \approx 0$ (if $Tc_2>Tc_1$, $Tc_2$ is then the critical temperature for the case of a very large size PNR). Value of $T_L$ is different for the case of QP and Vortex excitation. Depolarization field effect essentially decreases value of $T_L$ for the case of QP excitation, $QPT_L > VT_L$ for the case of small size PNR. Therefore, the QP condensation (local phase transition) is expected to occur before the Vortex condensation upon cooling. For the case of large size PNR , $QPT_L < VT_L$ . We also found that local TO modes, for which value of frequency $\omega_L < \omega_0(1)$, can exist due to reflection by the host medium. In the last part of our paper, we calculate QP scattering by the PNR taking into account depolarizing field effects and find strong resonance scattering at small TO momentum as described above. All calculations are exact as we do not assume that the TO–TO interaction between PNR and host medium is small. A short discussion of the results and future perspective are presented in the conclusion section.

## 2. MODEL

### *2.1 MODEL DESCRIPTION.*

[2] Similar photon excitations are called as magnetic and electric in electrodynamics [16].

We consider the dynamics of a spherically symmetric PNR embedded in an isotropic medium [14, 15]. The total Lagrangian density $L$ contains the contributions of the TO and LO waves, $L_o$, the electrostatic field, $L_{em}$, and the interaction between electric field and polarization, $L_{int}$. [23, 24].

The total Lagrangian $\Lambda$ is written as:

$$\Lambda=\int L d^3 r, \quad L=L_o+L_{em}+L_{\text{int}} \qquad (1)$$

$$L_o = K_o - \Pi_o \quad \text{where } K_o = 1/2\rho_o(\partial\xi/\partial t)^2 \text{ and } \Pi_o = +(A_o/2)div(\xi)^2 + B_o\tilde{\xi}_{ik}^{\,2} + 1/2\rho_o\omega_0^2\xi^2$$

$$\xi_{ik} = 1/2(\partial\xi_i/\partial x_k + \partial\xi_k/\partial x_i), \quad \tilde{\xi}_{ik} = \xi_{ik} - 1/3\delta_{ik}\xi_{qq} \qquad (2)$$

$$L_{em} = \frac{1}{8\pi}(\nabla\varphi)^2 \qquad (3)$$

$$L_{\text{int}} = -e^*\xi_\alpha\nabla_\alpha\varphi \qquad (4)$$

Here $\boldsymbol{\xi}$ is the displacement in the optic mode, $\rho_o$ the corresponding density, $e^*$ the effective charge. $A_o$, and $B_o$ are parameters describing the optic mode contribution to the potential energy written in a similar form to that in elasticity theory [25] and $\varphi$ is the electric scalar potential. A similar Lagrangian was proposed by Hopfield [26] for the description of polariton dynamics. Our Lagrangian is different from that proposed by Hopfield in three aspects. First, we do not take into account polariton effects. Second, we take into account the space dispersion of optical modes (see $\Pi_o$ (1) first and second terms), and, third, we also take into account the softening of the optic mode in the last term of $\Pi_o$:

$$\omega_0^2(1) = a1^2(T-Tc_1), \quad \omega_0^2(2) = a2^2(T-Tc_2), \quad Tc_1 < Tc_2, \ \text{a1} < \text{a2} \qquad (5)$$

A phase transition is expected to occur inside the PNR at a higher temperature than in the host medium, $T_{c1}<T_{c2}$. We should also note that Hopfield used the expression $L_H$ for the description of the interaction

$$L_H = e^*\varphi\nabla_\alpha\xi_\alpha = L_{\text{int}} + e^*\nabla_\alpha(\xi_\alpha\varphi) \ .$$

between electric field and polarization,

Both expressions $L_H$ and (4) lead to the same results for the case of infinitely large medium. We prefer apply expression (4) which is similar to the standard description of interaction between polarization and electric field and polarization [1, 24] and more suitable for the boundary conditions formulation. Given the above Lagrangian, the dynamical equations are written in the standard form:

$$d/dt(\delta\Lambda/\delta(d\xi_\alpha(r)/dt)) = \delta\Lambda/\delta\xi_\alpha(r), \quad \delta\Lambda/\delta\varphi(r) = 0 \qquad (6)$$

From eq.2 we then get:

$$\rho_0\partial^2\xi_\alpha/\partial t^2 = Ao * grad_\alpha div(\xi) + 2Bo * \nabla_k(\tilde{\xi}_{\alpha k}) -$$
$$-\rho_0\omega_o^2\xi_\alpha - e^*\nabla_\alpha\varphi - Ao\int div(\xi)dS_\alpha - 2Bo*\int\tilde{\xi}_{\alpha k}dS_k \qquad (7a)$$

and from eqs.3 and 4:

$$-\nabla_\alpha(\nabla_\alpha\varphi - 4\pi e^*\xi_\alpha) + \int(\nabla_\alpha\varphi - 4\pi e^*\xi_\alpha)dS_\alpha = 0 \qquad (7b)$$

The electrodynamics boundary condition for the tangential component of the electric field is given by:

$$\nabla\varphi_1 - \mathbf{n}(\mathbf{n}\nabla\varphi_1) = \nabla\varphi_2 - \mathbf{n}(\mathbf{n}\nabla\varphi_2) \qquad (7c)$$

where ***dS*** is an element of the PNR surface and ***n*** is a unit vector normal to the surface. The terms containing ***dS*** (7a, b) lead to particular solutions satisfying the boundary conditions at the PNR interface. (7b, c) are the standard boundary conditions of electrodynamics. It is important to emphasize that all parameters are the same inside and outside the PNR with the exception of the gaps, $\boldsymbol{\omega_0(1)}$ and $\boldsymbol{\omega_0(2)}$, determined by the transition temperatures $T_{c1}$ and $T_{c2}$ respectively.
We now consider separate independent solutions inside and outside the PNR in the form of plane waves with frequency $\omega$ and momentum $\boldsymbol{K_L}$, $\boldsymbol{K_T}$ for the optic longitudinal and transverse modes respectively. From eqs.7a and 7b:
*Longitudinal Optic wave (LOW)*

$$[\rho_o\omega^2 - AoK_L^2 - 4/3BoK_L^2 - \rho_o\omega_0^2]\xi_L - iK_L e^*\varphi = 0,$$
$$iK_L\varphi - 4\pi e^*\xi_L = 0 \qquad (8a)$$

Solving this system of two coupled equations, we get:

$$\omega^2 - \boldsymbol{AoK_L}^2/\rho_{\boldsymbol{o}} - 4/3\boldsymbol{BoK_L}^2/\rho_{\boldsymbol{o}} - \omega_0^2 - \Omega_{\boldsymbol{p}}^2 = 0 \text{ with } \Omega_{\boldsymbol{p}}^2 = 4\pi(\boldsymbol{e}^*)^2/\rho_{\boldsymbol{o}} \qquad (8b)$$

Because the ion plasma frequency, $\boldsymbol{\Omega_p}$~15 - 75 meV >>$\omega_o$, is large, the softening of the LOW can be neglected in practice.

*Transverse Optic wave (TOW)*

$$[\omega^2 - BoK_T^2/\rho_o - \omega_0^2]\xi_T = 0 \qquad (9)$$

We should note that combination of the longitudinal (8a,8b) and transverse (9) optic mode *isn't enough* for the description of the full solution of the dynamical eq-ns , depolarization field and corresponding displacement mode, DM, should be taken into account. DM space dependence
(~$\boldsymbol{r^{j-1}}$ inside PNR and ~$\boldsymbol{r}^{-j(j+1)-1}$ outside of PNR) is quite different from ~sin($\boldsymbol{qr}$-$\boldsymbol{\omega t}$) corresponding plane TO wave. DF, DM features will be discussed more detailed in the section 2.3 of this paper.
For convenience, we write the parameter $\boldsymbol{A_0}$ and $\boldsymbol{B_0}$ in the LOW and TOW dispersion, Eqs. 8b and 9, in terms of the speeds $C_L$, $C_O$:

$$Ao \equiv \rho_o[C_L^2 - 4/3\, C_O^2],\quad Bo \equiv \rho_0 C_O^2,\ \xi \to \xi/\sqrt{\rho_o} \qquad (10)$$

We also take the density parameter, $\boldsymbol{\rho_0}$, to be equal to 1.

## *2.2 MODEL SOLUTION IN THE ANGULAR MOMENTUM REPRESENTATION.*

Because the effective potential energies inside or outside the spherical PNR are independent of the coordinates, the solution can be decomposed into independent radial and angular parts. The angular part is written in terms of spherical functions. The radial part can be written in terms of Bessel (Hankel) functions inside (outside) the PNR with argument $rK_T$ and $rK_L$ for the TOW and LOW respectively (8a, 8b, 9). We are especially interested in the dynamics of transverse excitations. A similar problem was considered in quantum electrodynamics [16]. Our case is similar but does not quite coincide with [16] because the longitudinal optic modes are important in our case while they are absent in [16], and because the optic mode energy spectrum contains a gap. At last we should take into account the presence of the depolarization fields and corresponding optics wave displacements. Because we are interested in scattering by a spherical PNR and the total angular momentum is conserved, it is convenient to use [14, 15] an angular momentum technique in the following. For convenience, we utilize some results from Refs. [16], [17] and [27] concerning spherical vector functions. We describe the optic phonon polarization in terms of the vector spherical functions, $\chi_\mu(\alpha)$, μ=0, ±1, α=1,2,3 (x, y, z):

$$\chi_{-1} = \frac{1}{\sqrt{2}}\begin{pmatrix}1\\-i\\0\end{pmatrix},\ \chi_0 = \begin{pmatrix}0\\0\\1\end{pmatrix},\ \chi_1 = \frac{-1}{\sqrt{2}}\begin{pmatrix}1\\i\\0\end{pmatrix} \tag{11}$$

The phonon wave field $Y_{jlM}$ contains angular and polarization ("spin") components in the *total* angular momentum ***j***
representation:

$$\boldsymbol{Y_{jlM}}(\mathbf{v},\alpha) = \sum_{\boldsymbol{m},\mu} \boldsymbol{C^{jM}_{lm1\mu}} \boldsymbol{Y_{lm}}(\mathbf{v})\chi_\mu(\alpha), \text{ with } \boldsymbol{m}+\mu=\boldsymbol{M} \text{ and } \mathbf{v}=\mathbf{k}/\boldsymbol{k},$$

$$\mathbf{v}^2 = 1,\quad \boldsymbol{l}=\boldsymbol{j},\boldsymbol{j}\pm 1 \text{ if } \boldsymbol{j}\neq 0;\ \boldsymbol{l}=1 \text{ if } \boldsymbol{j}=0 \text{ and } \boldsymbol{M}=\boldsymbol{-j...j} \tag{12}$$

where ***k*** is the momentum, $Y_{lm}$ is a standard spherical function [16, 17] with angular momentum ***l*** and its projection *m* and $C^{jM}_{lm1\mu}$ is the Clebsch-Gordon coefficient [16]. We should emphasize that we are working here in the momentum (not-space) representation. Expression (12) can be written in the form of orthogonal normalized spherical vector functions (integration is done over the surface of a sphere of unit radius).

$$\mathbf{Y}_{\boldsymbol{jlM}}(\mathbf{v}) = \sum_{\boldsymbol{m},\mu} \boldsymbol{C^{jM}_{lm1\mu}} \boldsymbol{Y_{lm}}(\mathbf{v})\chi_\mu,\quad \int \mathbf{Y}^*_{\boldsymbol{jlM}}(\mathbf{v})\mathbf{Y}_{\boldsymbol{j'l'M'}}(\mathbf{v})d\Omega = \delta_{\boldsymbol{jj'}}\delta_{\boldsymbol{ll'}}\delta_{\boldsymbol{MM'}} \tag{13}$$

We distinguish several solutions: one longitudinal, with spherical vector $Y^{(-1)}_{jM}(\mathbf{v})$ (superscript -1), directed along the momentum ***k***, and two transverse, $Y^{(0)}_{jM}(\mathbf{v})$ and $Y^{(1)}_{jM}(\mathbf{v})$ (superscript 0 and +1), with different parity). Several useful expressions concerning spherical wave functions [16] are used in the paper:

$$\mathbf{Y}^{(-1)}_{jM}(\mathbf{v}) \equiv \mathbf{v}Y_{jM}(\mathbf{v}),\ \mathbf{Y}^{(0)}_{jM}(\mathbf{v}) \equiv \mathbf{Y}_{jjM}(\mathbf{v}),\ \mathbf{Y}^{(1)}_{jM}(\mathbf{v}) \equiv i[\mathbf{Y}_{jjM}(\mathbf{v}),\mathbf{v}],$$
$$\mathbf{Y}^{(0)}_{jM}(\mathbf{v})^*\mathbf{v}=0,\ \mathbf{Y}^{(0)}_{jM}(\mathbf{v})=-i\frac{[\mathbf{k}\nabla_{\mathbf{k}}]Y_{jM}(\mathbf{v})}{\sqrt{j(j+1)}},$$
$$\mathbf{Y}^{(1)}_{jM}(\mathbf{v})*\mathbf{v}=0,\ \mathbf{Y}^{(1)}_{jM}(\mathbf{v}) = \frac{k}{\sqrt{j(j+1)}}\nabla_{\mathbf{k}}Y_{jM}(\mathbf{v}), \tag{14}$$

$$\mathbf{Y}^{(-1)}_{jM}(\mathbf{v}) = \sqrt{\frac{j}{2j+1}}\mathbf{Y}_{j,j-1,M}(\mathbf{v}) - \sqrt{\frac{j+1}{2j+1}}\mathbf{Y}_{j,j+1,M}(\mathbf{v}),$$
$$\mathbf{Y}^{(1)}_{jM}(\mathbf{v}) = \sqrt{\frac{j}{2j+1}}\mathbf{Y}_{j,j+1,M}(\mathbf{v}) + \sqrt{\frac{j+1}{2j+1}}\mathbf{Y}_{j,j-1,M}(\mathbf{v}) \tag{15}$$

We note that the functions $\boldsymbol{Y_{jM}^{(0)}(v)}$, $\boldsymbol{Y_{jM}^{(1)}(v)}$ correspond to Vortex (V) and Quasi polar (QP) excitations respectively. A Vortex excitation is a manifold of closed lines of optic mode displacements (similar to the magnetic field lines in electrodynamics), and therefore does not possess an electric dipole moment. A QP TOW displacement (15) contains a spatial component corresponding to the spherical symmetric *S*- *orbital* state but also spin angular momentum, giving a non-zero electric dipole moment after integration over angles when $j$=1. We remind also that only the longitudinal wave exists when the total angular momentum $j$=0.

Next, in order to fulfill the boundary conditions at the spherical PNR surface, the phonon wave field must be expressed in real space. This can be done by means of an expansion of the plane wave in terms of spherical harmonics [16]:

$$\exp(i\mathbf{kr}) = \sum_{l,m} g_l(kr) Y^*_{lm}(\mathbf{k}/k) Y_{lm}(\mathbf{n}), \quad \mathbf{n} = \mathbf{r}/r, \quad n^2 = 1,$$

$$g_l(kr) = (2\pi)^{3/2} i^l J_{l+1/2}(kr)/\sqrt{kr},$$

$$\int \mathbf{Y}_{jlM}(\mathbf{k}/k)\exp(i\mathbf{kr})\,d\Omega_k = g_l(kr)\mathbf{Y}_{jlm}(\mathbf{r}/r) \qquad (16)$$

$J_{l+1/2}$ is an ordinary Bessel function. As usual, Bessel functions are used in expression (16) inside PNR to ensure a finite TOW displacement at the PNR center, r=0; outside the PNR, we should use the first (second) order Hankel $H_{l+1/2}^{(1),(2)}$ function for the case of an "outward wave" ("inward wave") instead of a Bessel function. According to our earlier notation, the wave vector outside (inside) the PNR is labeled as $k=K_T(1)$ ($k=K_T(2)$). In the real space representation, the wave fields corresponding to the $Y^{(0)}_{jM}(\boldsymbol{v})$, $Y^{(1)}_{jM}(\boldsymbol{v})$, *and* $Y^{(-1)}_{jM}(\boldsymbol{v})$ components are transformed into the expressions $\boldsymbol{F}^{(0)}_{jM}$, $\boldsymbol{F}^{(1)}_{jM}$, *and* $\boldsymbol{F}^{(-1)}_{jM}$, respectively:

$$\mathbf{F}^{(0)}_{jM} = g_j(kr)\mathbf{Y}^{(0)}_{jM}(\mathbf{n}) \equiv f^{(0)}_j(kr)\mathbf{Y}^{(0)}_{jM}(\mathbf{n}), \qquad (17a)$$

$$\mathbf{F}^{(1)}_{jM} = \left[\frac{j}{2j+1} g_{j+1}(kr) + \frac{j+1}{2j+1} g_{j-1}(kr)\right]\mathbf{Y}^{(1)}_{jM}(\mathbf{n}) +$$
$$\frac{\sqrt{j(j+1)}}{2j+1}\left[-g_{j+1}(kr) + g_{j-1}(kr)\right]\mathbf{Y}^{(-1)}_{jM}(\mathbf{n}) \equiv$$
$$f^{(1,1)}_j(kr)\mathbf{Y}^{(1)}_{jM}(\mathbf{n}) + f^{(1,-1)}_j(kr)\mathbf{Y}^{(-1)}_{jM}(\mathbf{n}) \qquad (17b)$$

$$\mathbf{F}^{(-1)}_{jM} = \left[\frac{j}{2j+1} g_{j-1}(kr) + \frac{j+1}{2j+1} g_{j+1}(kr)\right]\mathbf{Y}^{(-1)}_{jM}(\mathbf{n}) +$$
$$\frac{\sqrt{j(j+1)}}{2j+1}\left[g_{j-1}(kr) - g_{j+1}(kr)\right]\mathbf{Y}^{(1)}_{jM}(\mathbf{n}) \equiv$$
$$f^{(-1,-1)}_j(kr)\mathbf{Y}^{(-1)}_{jM}(\mathbf{n}) + f^{(-1,1)}_j(kr)\mathbf{Y}^{(1)}_{jM}(\mathbf{n}) \qquad (17c)$$

where we have introduced the coefficients $f^{(0)}_j$, $f^{(1,1)}_j$, $f^{(1,-1)}_j$, $f^{(-1,1)}_j$, $f^{(-1,-1)}$ for the expansion of the wave field components in terms of the orthogonal basis functions $\boldsymbol{Y}^{(0)}_{jM}(\boldsymbol{n})$, $\boldsymbol{Y}^{(1)}_{jM}(\boldsymbol{n})$, $\boldsymbol{Y}^{(-1)}_{jM}(\boldsymbol{n})$.

It is important to note that, when going from the momentum to the real-space representation, a transverse vortex (0) excitations converts to a transverse vortex excitation while a transverse quasi-polar (+1) converts to a combination of transverse (+1) and longitudinal (-1 ) excitations (17b). Hence, the index ±1 in the space representation and ±1 in the momentum representation do not have the same physical meaning. Polarization is normally defined with respect to $\boldsymbol{k}$, i.e. in the momentum representation. However, the boundary conditions are applied in real space, so that the conversion from momentum to real space representation is necessary. The main physical characteristic of the TOW (div(ξ)=0) is of course invariant with respect to the type of representation used. The functions $\boldsymbol{F}^{(0)}_{jM}$, $\boldsymbol{F}^{(+1)}_{jM}$ are transverse in the momentum representation ($\mathbf{F}^{(0)}_{jM}\,\mathbf{v}=0$, $\mathbf{F}^{(+1)}_{jM}\,\mathbf{v}=0$), i.e. with respect to k. In the space presentation however, the function $\boldsymbol{F}^{(0)}_{jM}$ is still transverse but this time with respect to $\boldsymbol{r}$ and the function $\boldsymbol{F}^{(+1)}_{jM}$ contains a *longitudinal* $\boldsymbol{Y}^{(-1)}_{jM}\boldsymbol{r}\neq 0$ term (17b). This term describes near field effects and is proportional to $1/r^2$ in the wave zone (kr>>l). At a large distance from the center of the PNR,

$$kr \to \infty, \quad g_l(kr) \approx (2\pi)^{3/2}\sqrt{\frac{2}{\pi}}(i)^l \frac{\sin(kr-\frac{l\pi}{2})}{kr} \qquad (18)$$

We refer to $\boldsymbol{F}^{(0)}_{jM}$ as the "Vortex" and to $\boldsymbol{F}^{(1)}_{jM}$ as the "quasi polar", QP, components of the TOW. In a QP $\boldsymbol{F}^{(1)}_{jM}$ type excitation, both transverse and longitudinal optical mode components will be present as a result of applying the boundary conditions (7a), because these BCs are stated for the tangential and normal components of the displacements and stresses, with respect to $\boldsymbol{r}$ and not $\boldsymbol{k}$. By contrast, the Vortex $\boldsymbol{F}^{(0)}_{jM}$ excitation will correspond to a strictly transverse excitation and will not be mixed with a LOW, since it only contains a $\boldsymbol{Y}^{0}_{jM}$ term, as seen in expression (17a).
The general displacement $\boldsymbol{\xi}_L$ in the longitudinal optical wave (8b) can also be written as the gradient of some scalar "potential" $\Phi_o$ (19).

$$\xi_L = \nabla\Phi_o, \text{ with } \Phi_o = G_j(r)Y_{jM}(\vartheta,\phi) \qquad (19)$$

Writing ξ in terms of the potential $\phi_o$ in eqs. 7a and 7b, we get:

$$\Delta\Phi_o + (\omega^2 - \omega_0^2 - \Omega_p^2)/C_L^2\,\Phi_o = 0 \qquad (20a)$$

and

$$\frac{1}{r^2}\frac{\partial}{\partial r}\left(r^2\frac{\partial G_j(r)}{\partial r}\right) - \frac{j(j+1)}{r^2}G_j(r) + \frac{(\omega^2-\omega_0^2-\Omega_p^2)}{C_L^2}G_j(r) = 0 \quad (20b)$$

Function $\boldsymbol{G_j(r)}$ is proportional Bessel or Hankel functions from argument $\boldsymbol{K_L r}$. The LOW displacement (21) contains not only a component $\propto\boldsymbol{Y}^{(-1)}_{jM}(\boldsymbol{n})$ that is perpendicular to surface of the PNR but also a component $\propto\boldsymbol{Y}^{(1)}_{jM}(\boldsymbol{n})$ that is parallel to it ($\boldsymbol{n=r/r}$, $\boldsymbol{n}^2=1$).

$$\xi_L = \frac{\partial G_j}{\partial r}\mathbf{n}Y_{jM} + G_j\nabla Y_{jM} = \frac{\partial G_j}{\partial r}\mathbf{Y}^{(-1)}_{jM}(\mathbf{n}) + G_j\frac{\sqrt{j(j+1)}}{r}\mathbf{Y}^{(1)}_{jM}(\mathbf{n}) \qquad (21)$$

The near field effects described by the term $\boldsymbol{Y}^{(1)}_{jM}(\boldsymbol{n})G_j/r$ in eq.21 rapidly decay at a large distance $r$ from the PNR center $\sim 1/r^2$. Also, we are interested by the case when the ion plasma frequency in (8b), $\boldsymbol{\Omega_p}$, is large, $\Omega_p \gg \omega$. Such a LOW will only be excited within a ***narrow layer*** of thickness $\boldsymbol{\delta R}$ at the PNR surface, $\boldsymbol{\delta R}\sim 1/|2\pi \boldsymbol{K_L}|$ (8b). For the case of a relaxor crystal, a typical PNR radius is $\boldsymbol{R}\sim 4 \div 15$ l.u. and $\delta R\sim 1$ l.u. if $\boldsymbol{\Omega_p}$=30meV and $\boldsymbol{C_L}$=100 meV*l.u.
Let's consider depolarization field and corresponding depolarization mode displacement. We have the ***full*** solution of Eqn. (7b***) outside of PNR sur*****face** taking into account eq-s (19, 20a, 20b)

$$\Delta\varphi - 4\pi e^{*}(\nabla_\alpha \xi_\alpha) = 0,\ \boldsymbol{\xi}_L = \nabla\Phi_0,\ \Phi_0 = G_j(r)Y_{jM}(\vartheta,\phi),$$

$$\varphi = 4\pi e^{*}(\Phi_0 + \delta\tilde{\varphi}),\ \Delta\delta\tilde{\varphi} = 0,\ \delta\tilde{\varphi} \equiv \delta\tilde{\varphi}_j(\mathrm{r})Y_{jM}(\vartheta,\phi),$$

$$\frac{\partial}{\partial r}\left(r^2 \frac{\partial\delta\tilde{\varphi}_j(\mathrm{r})}{\partial r}\right) - j(j+1)\delta\tilde{\varphi}_j(\mathrm{r}) = 0 \qquad (22)$$

Depolarization field $\delta\tilde{\varphi}_j(r)$ which is limited inside and outside PNR is found from Eq-n (22):

$$\delta\tilde{\varphi}_j(r) = A_\delta^{(2)} r^j \theta(R-r) + A_{\delta,-1}^{(1)} r^{-j-1} \theta(r-R),\ \theta(x<0)=0,\ \theta(x>0)=1 \qquad (23)$$

where $\boldsymbol{A}_{\delta}^{(2)}$, $\boldsymbol{A}_{\delta,-1}^{(1)}$ –arbitrary constants and $\boldsymbol{R}$- PNR radius. DF leads to the additional OW displacement. Such depolarization mode (DM) displacements are calculated by the direct substitution DF $\delta\tilde{\varphi}$ (23) to the eq-n (7a):

$$\delta\xi_{L,\alpha}^{(2)} = \nabla_\alpha(C_{2D} r^j Y_{jM}),\ C_{2D} = \frac{\Omega_p^2}{(\omega^2 - \omega_o^2(2))} A_\delta^{(2)},\ r < R \qquad (24a)$$

$$\delta\xi_{L,\alpha}^{(1)} = \nabla_\alpha(C_{1D} r^{-(j+1)} Y_{jM}),\ C_{1D} = \frac{\Omega_p^2}{(\omega^2 - \omega_o^2(1))} A_{\delta,-1}^{(1)},\ r > R \qquad (24b)$$

We note that expressions (24a, 24b) are independent vs. parameters ***Ao, Bo*** (7a) defined OW space dispersion. DM displacements are oscillated over time but its space dependence ($\sim \boldsymbol{r}^{j-1}$ inside PNR and $\sim \boldsymbol{r}^{-j(j+1)-1}$) outside of PNR) is quite different from the expressions ~sin($\boldsymbol{qr}$-$\boldsymbol{\omega t}$), cos($\boldsymbol{qr}$-$\boldsymbol{\omega t}$) corresponding plane TO wave. DM displacements create no charge density fluctuations due to the existence of the equalities $\nabla_\alpha \delta\xi_{L,\alpha}^{(2)} = \nabla_\alpha \delta\xi_{L,\alpha}^{(1)} = 0$. Simultaneously DM displacements are longitudinal in a meaning $\mathbf{curl}(\delta\boldsymbol{\xi}_L^{(2)}) = \mathbf{curl}(\delta\boldsymbol{\xi}_L^{(1)}) = 0$. DM displacement corresponding ***j***=1 is depended vs. spherical angles but independent over the value of radius ***r*** inside of PNR. The electrostatic potential $\boldsymbol{\varphi}$ is included in Eqn. (7a) in the form $\partial\phi/\partial r_\alpha$ and therefore doesn't affect the TOW dynamics. Due to the equality $\Delta\delta\phi=0$, the electrostatic depolarizing potential component, $\delta\phi$, doesn't affect the LOW dynamics (8b).
DM displacements are excited due to the scattering of the “ordinary” excitations at the boundaries.
We now use the general solution of the model to describe the scattering of an incident transverse optic (TO) plane wave by a spherical PNR. Expanding the plane wave in terms of spherical harmonics [17], we obtain an expression for the partial wave $\boldsymbol{\sigma}^{(0)}_j$, $\boldsymbol{\sigma}^{(1)}_j$ and total, $\boldsymbol{\sigma}^{(0)}$, $\boldsymbol{\sigma}^{(1)}$ scattering cross-section:

$$\sigma^{(0,1)} = \sum_{j\geq 1} (2\mathrm{j}+1)\sin(\delta_{\mathrm{j}}^{(0,1)})^2\, 2\pi/k^2 \qquad (25)$$

$\delta_j^{(0)}$ and $\delta_j^{(1)}$ are the phase shift for scattering of vortex and quasi-polar excitations respectively.

## *2.3 BOUNDARY CONDITIONS FOR THE CASE OF VORTEX EXCITATIONS*

TOW Vortex excitations create no charge density fluctuations and no mixing with LOW at the PNR surface. Depolarization fields are absent in this case. Therefore boundary conditions for the case of Vortex excitations are the same which were found early [14, 15]:

$$\Psi(g) = 0, \quad \frac{\partial \Psi(g)}{\partial r} = 0 \tag{26a}$$

where the auxiliary function Ψ(g) represents the difference between the TO amplitudes outside and inside the PNR:

$$\Psi(g) = A^{(0)}_{in} g_{j,in} + A^{(0)}_{\text{out}} g_{j,out} - A^{(0)}(2) g_j \tag{26b}$$

$A^{(0)}(2)$, $A^{(0)}_{in}$, $A^{(0)}_{out}$ are the Vortex amplitudes respectively inside the PNR, incident (in) and reflected (out) from the PNR surface on the outside; $g_j$, $g_{j,out}$, $g_{j,in}$ – spherical Bessel function, first and second order spherical Hankel function, respectively. Eq-ns (26a) describe continuing of the Vortex displacement and it derivative over radius at the PNR surface.

### *2.4 BOUNDARY CONDITIONS FOR THE CASE OF QUASI POLAR EXCITATIONS*

Electrostatic boundary condition (7c) including expression

$$\nabla_\alpha \varphi - \mathbf{n}_\alpha(\mathbf{n}_\beta \nabla_\beta \varphi) = 4\pi e^* \frac{\sqrt{j(j+1}}{r} (G_j + \delta\tilde{\varphi}_j) \mathbf{Y}_{jM\alpha}{}^{(1)}$$

so that the value $\boldsymbol{G_j+\delta\tilde{\varphi}_j}$ (22, 23, 24a, 24b) should be continuing at the PNR surface, that is:

$$A_{2L} G_j^{(2)}(R) + C_{2D}(\omega^2 - \omega_0^2(2)) / \Omega_p^2 R^j =$$
$$A_{1L} G_j^{(1)}(R) + C_{1D}(\omega^2 - \omega_0^2(1)) / \Omega_p^2 R^{-j-1} \tag{27}$$

$\boldsymbol{A_{2L}}$ –LO amplitude inside PNR, $\boldsymbol{A_{1L}}$ – LO out wave, ~exp(I*KL*r), amplitude outside of PNR, we remind that corresponding momentum $\boldsymbol{K_L=I*\Omega_p/C_L}$ is the imaginary value.

Let's consider boundary condition DS=0 from the eq-n (7b),

$$DS \equiv \int (\nabla_\alpha \varphi - 4\pi e^* \xi_\alpha) dS_\alpha \tag{28a}$$

LOW (8a) satisfies to the boundary condition DS=0.

TOW doesn't excite electrostatic potential; TOW contribution to the value of DS at the internal part of the PNR surface is following:

$$\int -4\pi e^* \xi_\alpha dS_\alpha = -4\pi e^* A_2 f^{(2)}_{j,-1}$$

where $\boldsymbol{A_2}$ is the TOW amplitude and $\boldsymbol{f^{(2)}_{j,-1}}$ is the value of $\boldsymbol{f^{(1,-1)}_j}$ calculated for the region inside PNR. Depolarization field and corresponding displacement (24a, 24b) contribution to the value of DS at the internal part of the PNR surface is following:

$$4\pi e^* jr^{j-1} (A^{(2)}_\delta - C_{2D}) = 4\pi e^* jr^{j-1} \left(\frac{(\omega^2 - \omega_0^2(2))}{\Omega_p^2} - 1\right) C_{2D}$$

The similar calculation can be done for the respect of contribution to the value of DS at the external part of the PNR surface. Finally eq-n DS=0 could be written as following:

$$-A_2 f^{(2)}_{j,-1} + jr^{j-1}\left(\frac{(\omega^2-\omega_0^2(2))}{\Omega_p^2}-1\right)C_{2D} =$$
$$-A_{1in} f^{(1)}_{j,in,-1} - A_{1out} f^{(1)}_{j,out,-1} + (-j-1)r^{-j-2}\left(\frac{(\omega^2-\omega_0^2(1))}{\Omega_p^2}-1\right)C_{1D} \quad (28b)$$

where $\boldsymbol{A_{1in}}$, $\boldsymbol{A_{1out}}$ is the TOW in and out amplitude and $\boldsymbol{f^{(1)}_{j,in,-1}}$ ($\boldsymbol{f^{(1)}_{j,out,-1}}$) is the same as $\boldsymbol{f^{(1,-1)}_j}$ calculated for the in(out) TOW and outside PNR.
At last we have two boundary conditions corresponding continuing OW displacement perpendicular and parallel PNR radius and two boundary conditions corresponding continuing surface forces $\boldsymbol{\sigma_{\alpha\beta}n_\beta}$ (7a).
It is convenient to express displacements as the form

$$\boldsymbol{\xi} = G^{(1)}(r)\mathbf{Y}^{(1)}_{jM} + G^{(-1)}(r)\mathbf{Y}^{(-1)}_{jM} \quad (29)$$

For the case of TOW (we omit factors $A_2$, $A_{1in}$…)

$$G^{(1)}=G^{(1)}_T=\left[\frac{j}{2j+1}g_{j+1}(kr)+\frac{j+1}{2j+1}g_{j-1}(kr)\right]=f^{(1,1)}_j(\mathrm{kr}),$$
$$G^{(-1)}=G^{(-1)}_T=\frac{\sqrt{j(j+1)}}{2j+1}\left[-g_{j+1}(kr)+g_{j-1}(kr)\right]=f^{(1,-1)}_j(\mathrm{kr}) \quad (30a)$$

For the case of the LOW:

$$G^{(1)}=G^{(1)}_L=G_j\frac{\sqrt{j(j+1)}}{r},\quad G^{(-1)}=G^{(-1)}_L=\partial G_j/\partial r \quad (30b)$$

For the case of the depolarizing displacements, we have:

$$G_j(r<R)=r^j,\ G^{(1)}=\sqrt{j(j+1)}r^{j-1},\ G^{(-1)}=jr^{j-1},$$
$$G_j(r>R)=r^{-(j+1)},\ G^{(1)}=\sqrt{j(j+1)}r^{-j-2},\ G^{(-1)}=-(j+1)r^{-j-2} \quad (30c)$$

The boundary conditions (7a) is formulated as a continuity requirement on the surface force value $\sigma_{\alpha\beta}n_\beta$, that is continuing of the values of the force components $\Sigma_1$ and $\Sigma_{-1}$ perpendicular and parallel to the PNR radius respectively.

$$\sigma_{\alpha\beta} n_\beta = \mathbf{Y}^{(1)}_{jM,\alpha}\Sigma_1 + \mathbf{Y}^{(-1)}_{jM,\alpha}\Sigma_{-1},$$

$$\Sigma_1 = C_T^2 \left[\frac{\partial G^{(1)}}{\partial r} + \frac{\sqrt{j(j+1)}}{r} G^{(-1)} - \frac{1}{r} G^{(1)}\right],$$

$$\Sigma_{-1} = C_L^2 \left[\frac{\partial G^{(-1)}}{\partial r} - \frac{\sqrt{j(j+1)}}{r} G^{(1)} + \frac{2}{r} G^{(-1)}\right] + C_T^2 \left[\frac{2\sqrt{j(j+1)}}{r} G^{(1)} - \frac{4}{r} G^{(-1)}\right] \quad (31)$$

TOW contribution to $\Sigma_1$ and $\Sigma_{-1}$:

$$\Sigma_{1T} \equiv C_T^2 \left[\frac{\partial G_T^{(1)}}{\partial r} + \frac{\sqrt{j(j+1)}}{r} G_T^{(-1)} - \frac{1}{r} G_T^{(1)}\right]$$

$$\Sigma_{-1T} \equiv C_T^2 \left[\frac{2\sqrt{j(j+1)}}{r} G_T^{(1)} - \frac{4}{r} G_T^{(-1)}\right] \quad (32a)$$

LOW contribution to $\Sigma_1$ and $\Sigma_{-1}$ :

$$\Sigma_{1L} = C_T^2 \left[\frac{\partial G^{(1)}}{\partial r} + \frac{\sqrt{j(j+1)}}{r} G^{(-1)} - \frac{1}{r} G^{(1)}\right],$$

$$\Sigma_{-1L} = C_L^2 \left[\frac{\partial G^{(-1)}}{\partial r} - \frac{\sqrt{j(j+1)}}{r} G^{(1)} + \frac{2}{r} G^{(-1)}\right] + C_T^2 \left[\frac{2\sqrt{j(j+1)}}{r} G^{(1)} - \frac{4}{r} G^{(-1)}\right] \quad (32b)$$

Depolarization mode displacement contribution to $\Sigma_1$ and $\Sigma_{-1}$:

$$r < R,\ \Sigma_{1DM}(2) = C_T^2\, 2(j-1)\sqrt{j(j+1)}\, r^{j-2},$$

$$r < R,\ \Sigma_{-1DM}(2) = C_T^2\, 2j(j-1)\, r^{j-2},$$

$$r > R,\ \Sigma_{1DM}(1) = -C_T^2\, 2(j+2)\sqrt{j(j+1)}\, r^{-j-3},$$

$$r > R,\ \Sigma_{-1DM}(1) \equiv C_T^2\, 2(j+1)(j+2)\, r^{-j-3} \quad (32c)$$

Six eq-ns described boundary conditions allow to calculate scattering cross section, space distribution of TO, LO and depolarization mode displacements and so on. We will not demonstrate here these no complicated but cumbersome expressions but prefer directly show results of calculations.

## 3. RESULTS of CALCULATIONS

### *3.1. SOFT-MODE - TEMPERATURE DIAGRAM.*

We first plot in figure 1 borrowing from the ref. [15], the soft mode frequency, $\omega(T)$, inside and outside the PNR for the particular values of the transition temperatures, $T_{c2}$ and $T_{c1}$, used in the calculations below. General conclusions about the TOW dynamics are then obtained in the following subsections.

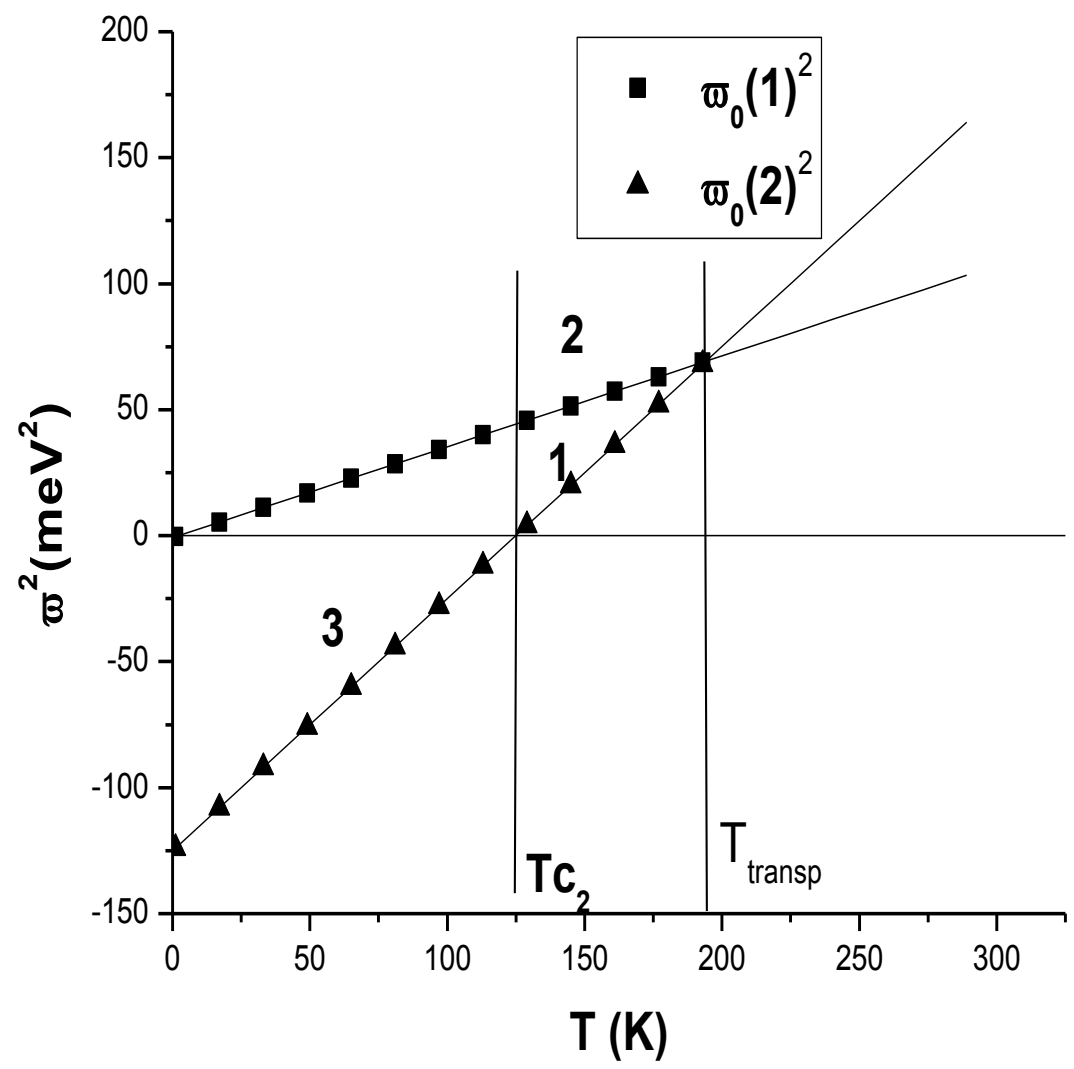

Fig.1. Soft mode energy gap versus temperature for the case of a large radius PNR. $\boldsymbol{Tc_1}$=2K, $\boldsymbol{Tc_2}$=125 K, $a1$=0.6; $a2$=1; $C_O$=70. $\omega_o(1)^2=a1^2(T\text{-}Tc_1)$, $\omega_o(2)^2=a2^2(T\text{-}Tc_2)$. Numerical values of parameters were estimated for the particular physical systems considered and are being used in following calculations. $T_{transp}$. is the temperature at which the soft-mode gaps inside and outside the PNR are the same and the soft-mode not scattered by the PNR. $T_{transp}=(a1^2Tc_1\text{-}a2^2Tc_2)/(a1^2\text{-}a2^2)$, $T_{transp}\approx 194.18$ *K*. 1 – triangle region where localized modes are possible; 2 – region where an incident TOW is scattered by PNR; 3 –condensed PNR phase.

Fig.1 couldn't be considered as general relaxor phase-diagram, it is rather model diagram corresponding to the simple, starting approximation. I limited himself to the temperature region $T<T_{transp}$ (for the simplicity).

3.2. *SYSTEM INSTABILITY DUE TO THE VORTEX AND QUASIPOLAR EXCITATIONS*

One PNR embedded in a very large host medium will be self transformed to another phase in process of local phase transition. It is assumed that temperature $\boldsymbol{T}>\boldsymbol{Tc_1}$ so that the host medium is in the high-temperature phase. A large PNR would undergo a phase transition at $\boldsymbol{T}=\boldsymbol{T_{c2}}$ (Fig.1). A finite size PNR will transform at a lower temperature $\boldsymbol{T}=\boldsymbol{T_{inst}}<\boldsymbol{Tc_2}$. We calculated the value of temperature instability, $\boldsymbol{T_{inst}}$, by the following method. The instability occurs when the frequency ω of the PNR TOW mode goes to zero, that is we self limit by the case of "second order local phase transition". In general our calculations are similar to ref. [14, 15], however we took in account depolarization field effect. Therefore, we should find the non-zero solutions of boundary conditions corresponding to both $\boldsymbol{\omega}$=0 and $\boldsymbol{A_{T1in}}$=0. The boundary conditions are written in the form of a matrix equation: $MA = 0$, where the components of the vector $\boldsymbol{A}$ are the amplitudes $\boldsymbol{A_{T2}}$, $\boldsymbol{A_{T1out}}$, $\boldsymbol{A_{L2}}$, $\boldsymbol{A_{L1out}}$, ***C2D, C1D*** and the 6x6 matrix $\boldsymbol{M}$ depends on two parameters – temperature, **T**, and PNR radius, $\boldsymbol{R}$. Eq-n det($\boldsymbol{M}$)=0 could have several solutions *Tinst* for both vortex and QP excitations, but only the first ones, corresponding to the highest temperature, are plotted in Fig. 2.

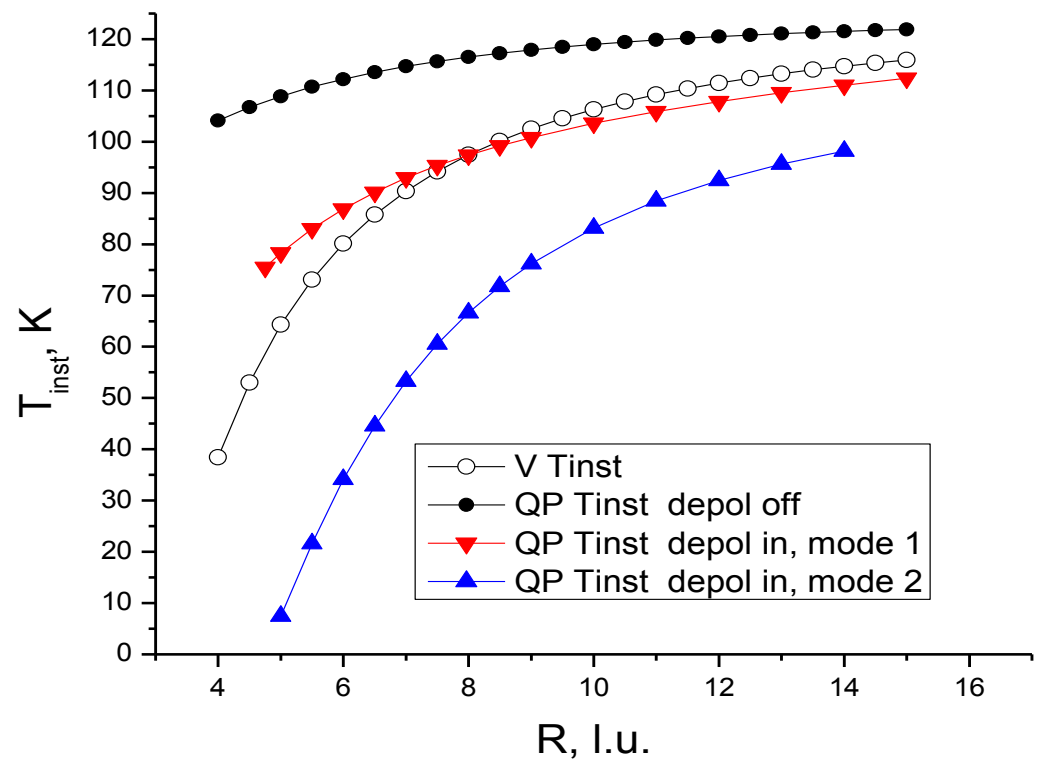

Fig.2. Temperature of the local instability for the case of Vortex, V $T_{inst}$, and Quasi Polar, QP $T_{inst}$ , excitations (depolarization field "in" and "off"), respectively, vs PNR radius *R*. Total angular momentum *j*=1. Values of parameters: $\boldsymbol{Tc_1}$=125 K, $\boldsymbol{Tc_2}$=2 K, ***a1***=0.6 meV/K, ***a2***=1 meV/K, $\boldsymbol{\Omega_p}$=20 mV, $\boldsymbol{C_o}$=70 meV/r.l.u., $\boldsymbol{C_L}$= 100 meV/r.l.u. These values of parameters will be used (as a rule) in the rest part of paper. Two QP instability mode contributions are shown for the depolarization field in case.

Temperature of instability is smaller for larger total angular momentum *j*> 1 and smaller PNR radius ***R***. This phenomenon is explained as in [15] by the presence of centrifugal barrier. Depolarization field effect is absent for the case of Vortex excitation. Depolarization field leads to the strong decreasing of the value of $\boldsymbol{T_{inst}}$ for the case of QP excitations. It is explained by following. DF is directed against the direction of the electric dipoles in the simple cases [18] and therefore leads to the increasing of the free energy and decreasing of the value of $\boldsymbol{T_{inst}}$.

Recently it was shown by the molecular dynamic calculations [20 – 22] that DF contribution to the free energy can lead to the complicated structure of the ground state, instead of simple ferroelectric we could expect for example the presence of the toroidal momentum of polarization.

Suppose that some PNR with radius ***R*** appears in process of sample cooling. If DF effect is suppressed due to the presence of the impurity charges and so on reasons then QP condensation will appears in the process of further cooling [15]. Vortex condensation could exist at the lower temperature, $VT_{instt}$, $<QPT_{inst}$ . Probably it is not contradicted to the most experimental data – PNR is described as polarized pieces of crystal. This sequence of events will be change for the case of non-screened DF: Vortex condensation will appears in the process of cooling (Fig.2) if PNR radius ***R***>8 l.u. QP condensation could exist at the lower temperature, $VT_{instt}$, $> QPT_{inst}$ . Authors' publication [20 – 22] supposed to observe Vortex condensation in the quantum dots. Unfortunately the definite experimental evidence concerning this phenomenon is still absent.

It seems reasonably to note some indirect evidence of the Vortex condensation in KTN relaxor ($KTa_{1-x}Nb_xO_3$, x=0.1÷0.15). Several experimental attempts undertaken at NIST and LLB by J.Toulouse et al (J. Toulouse, E. Iolin, B. Hennion, D. Petitgrand, R. Erwin, G. Yong, unpublished) to observe external electric field effect at the neutron scattering by the transversal acoustic phonons in KTN weren't successful, electric field effect was absent. However Xu et al [29] observed strong effect of the electric field at TAW scattering at PZN-4.55% PT relaxor which is explained by reorientation ordering PNR by the electric field. It seems reasonably to suppose that effect of depolarizing field is more strongly suppressed (screened) in chemical and valence inhomogeneous PZN-4.55 PT than in valence and ion radius homogeneous "weak relaxor" KTN. The effect of the electric field will be absent in KTN if vortex condensation appears in this crystal.

Decreasing of temperature leads to the appearing of the new mode instability (Fig.2). Number of these modes is increased for the case of large PNR size. For example mode instability appears at ***T***=112.3, 100.2, 82.7, 52.1, and 11.9 K for the case ***R***=15.

A quasi-polar excitations show the existence of a non-zero electric dipole moment in the spherical layers but only for *j*=1. Following the previous Eqs (17b) and (29, 30a, 30b and 30c), the average dipole moment density, ***P***, associated with the QP excitation can be written:

$$\mathbf{P}_{\mu} \equiv \int \mathbf{dO}\xi_{\mu} = \chi_{\mu}\{\mathbf{A}_{1out}\left(\sqrt{\frac{2}{3}}\mathbf{G}^{(1)} + \sqrt{\frac{1}{3}}\mathbf{G}^{(-1)}\right) + \cdots\} \quad (33)$$

Type equation here.

Here dO is spherical angles. The unit vector $\chi_{\mu}$ (11) defines the direction of the dipole moment ***P*** and is fixed extrinsically by interaction with chemical short range ordering, dislocations and so on. We note that depolarization displacement contribution to the average dipole moment ***P*** isn't depended vs *r* inside PNR (30c) and is equal to zero outside of PNR.

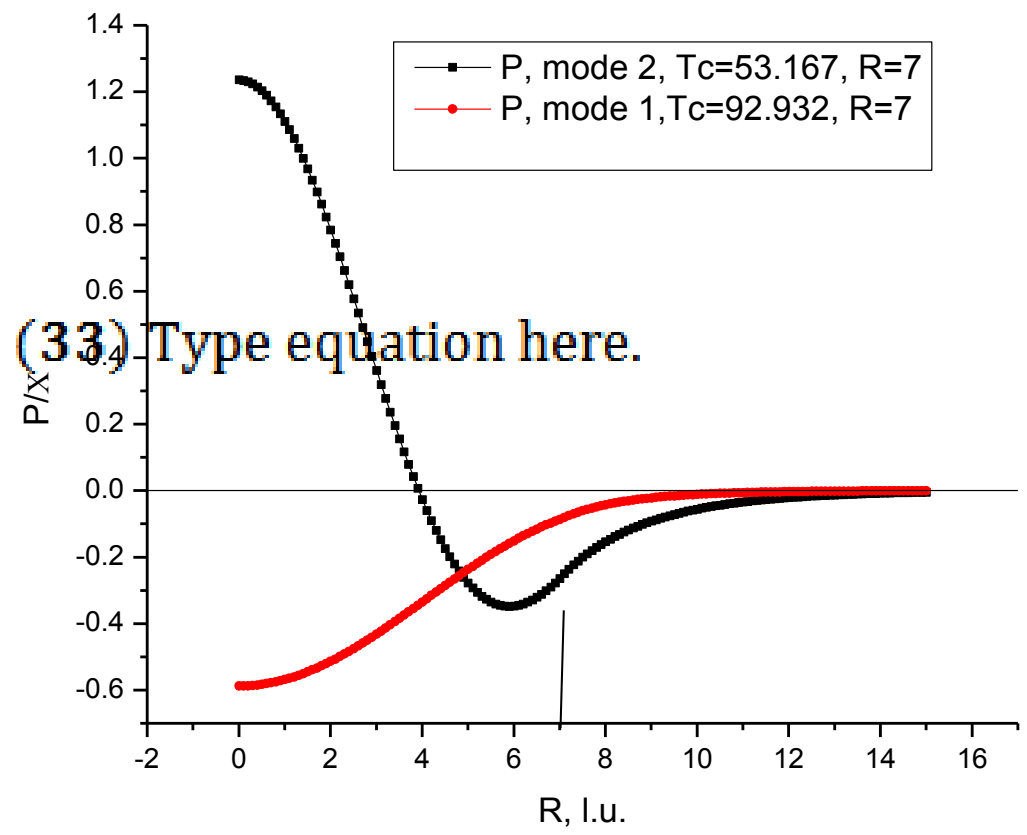

Fig.3 Radial layer distribution of the QP average dipole moment ***P*** near the temperature of the corresponding local phase transition: mode 1, $QPT_{inst}$=92.932K; mode 2, $QPT_{inst}$=53.167 K (critical modes, frequency ω=0). PNR radius ***R***=7, ***j***=1.

Dipole moment ***P*** is continuous function (in contrast to the case of suppressed depolarizing field and very large LOW frequency, $\Omega_p \rightarrow \infty$ [14, 15]). It is interesting that for the case of mode 2 value of ***P*** has different sign at the PNR center and side (also in contrast with [14, 15] result).

### 3.3 *LOCALIZED OPTIC MODES*

Local optic modes (LM) are characterized by an infinite lifetime as determined by the condition Im(ω)=0. They are in general discrete and exist in the "cage" created by the host media; the LM frequency is lower than the frequency of modes propagating outside the PNR, $\omega_{LM} < \omega_0(1)$. Therefore, LMs cannot be "directly" excited by a propagating incident TO wave and only shallow LMs can scatter it effectively. Quasi localized modes, QLM, have an energy that is greater than the soft mode energy gap of the host medium. They result from the effective long-time interaction between an incident TOW and the PNR. They are seen as higher amplitude of the propagating TO mode in the vicinity of the PNR and can lead to the resonant scattering. We limit ourselves here by the analysis of the LM. Calculations were done similar (in general) to the previously described [14, 15].

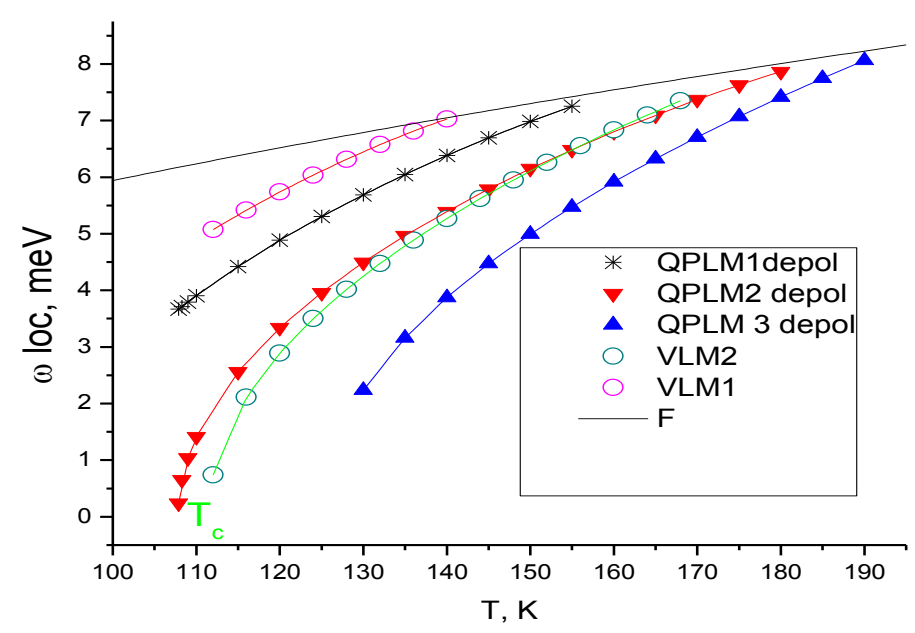

Fig.4. Frequency ω of the Vortex and QP localized excitations, LM; j=1. Several high and low frequency branches are shown as a function of temperature. The PNR radius R =12, that is Vortex condensation appears at the temperature T≈110K before QP condensation in process of cooling (Fig.2). Also shown is the temperature dependence of the host medium soft mode gap, F, which sets an upper limit for the value of the LM frequency.

Depolarization field has no effect at the Vortex local modes. However depolarizing field increases frequency of the QP LM and even creates new LM branch. General interpretation of these data is similar to the described in [14, 15]. Vortex and QP LMs could be studied by means of inelastic neutron scattering (in general).

### *3.4. TOW SCATTERING BY THE POLARIZED NANO REGION: STRONG EFFECT OF THE DEPOLARIZING MODES.*

The scattering cross-sections of transverse QP and V excitations have been calculated using eq. (25). Depolarizing modes aren't excited in process of Vortex scattering, corresponding V scattering cross section, $\Sigma_V$, was calculated and analyzing before [14,15]. This V-scattering demonstrates the presence of the strong long wave resonances explained by the existence of the quasi-localized states [14, 15]. For the case of very small incident wave momentum $\boldsymbol{q_1}$, the Vortex dynamics is similar to the scattering of a slow particle in quantum mechanics [17] and partial cross section of scattering, $\sigma_V \sim q_1^4$ for $q_1 \le 0.01$ and $j$=1. The similar results were found for the case of QP scattering in approximation where depolarizing field effect was omitted, $\sigma_{QP} \sim q_1^4$ [14, 15].

Here we calculate TOW QP scattering taking into account depolarizing field, DF, effect (see fig. 5-10). We found that *$\sigma_{QP}(j=1)$ isn't nullified at very small momentum $q_1$* , say *q1*~0.001, and nullified *at the extremely small value of momentum, q1*~1E-6 ÷ 1E-7 after passing of the extremely narrow peak of scattering.

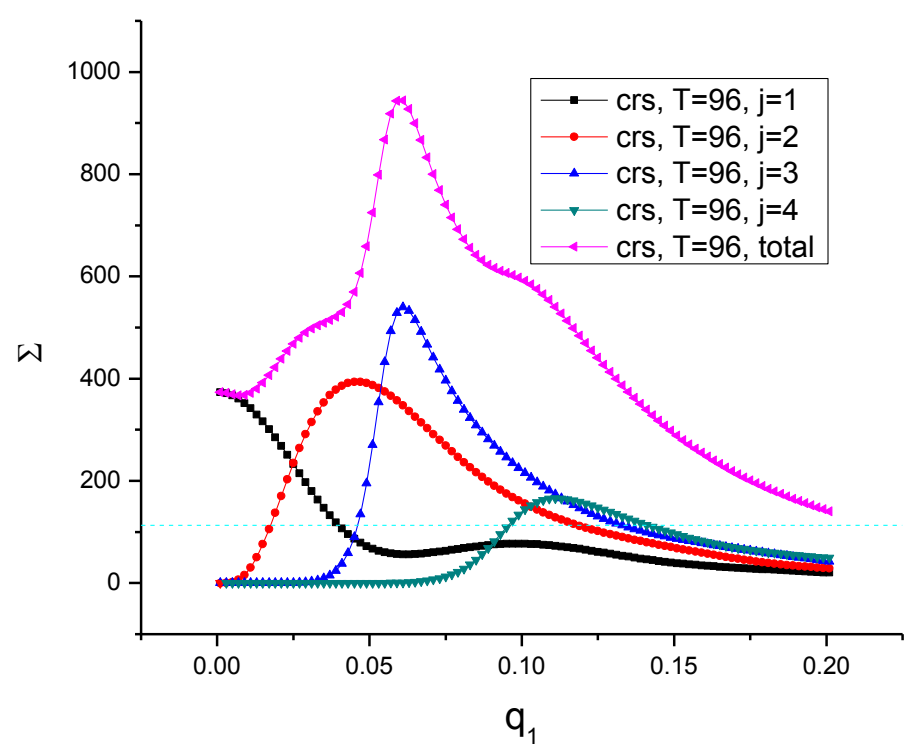

a)

Fig.5a. QP scattering by PNR (total and partial section Σ). The incident TOW momentum $\boldsymbol{q_1}$ is expressed in units π/**a,** parameter **a** is similar to the lattice constant; *R*=6, *T*=96K. The horizontal dash line marks the PNR geometrical cross-section.

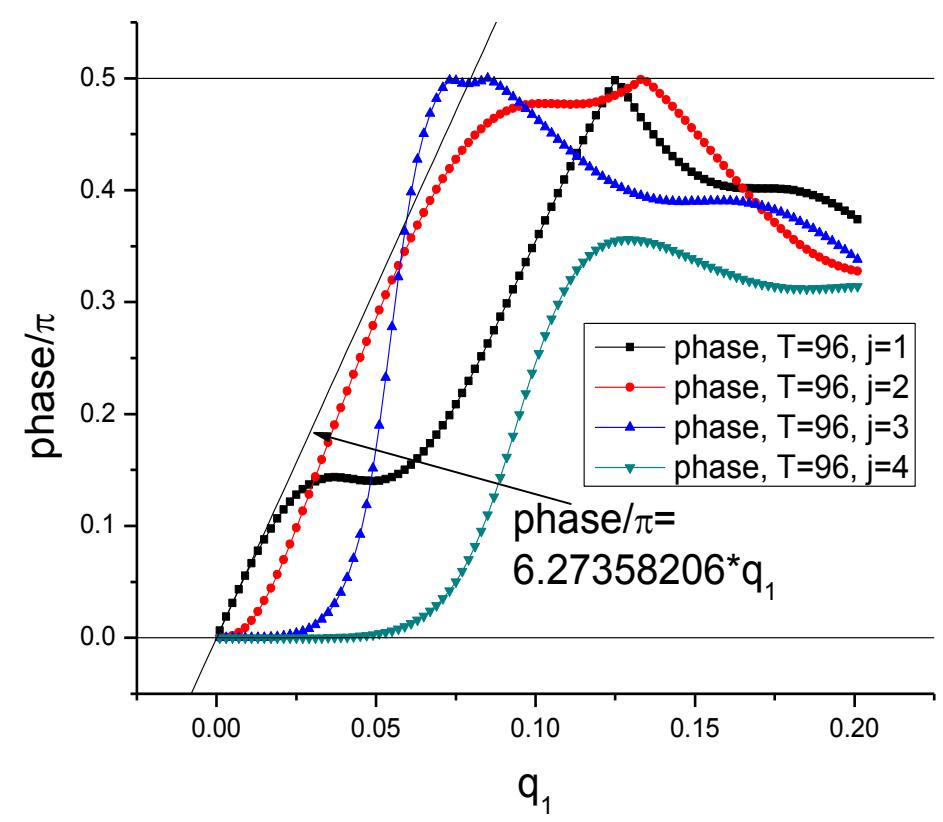

b)

Fig.5b. Phase of QP scattering by PNR. *R*=6, *T*=96K. The resonance appears when phase ≈ π/2. Position of the cross section maximum isn't coincided with resonance one due to the presence of a factor $1/\boldsymbol{q_1^2}$ in the Eq. (25).

We should underline unusually dependence phase of scattering at ***j***=1 vs $\boldsymbol{q_1}$, phase ~ $\boldsymbol{q_1^1}$ at the very small value of momentum $q_1$. This circumstance leads to the existence of plateau at the cross section ***σ(j=1)*** at the small momentum of the incident TOW (Fig.5a). Similar plateau is absent at the QP scattering at j≥2 and Vortex scattering at j≥1.

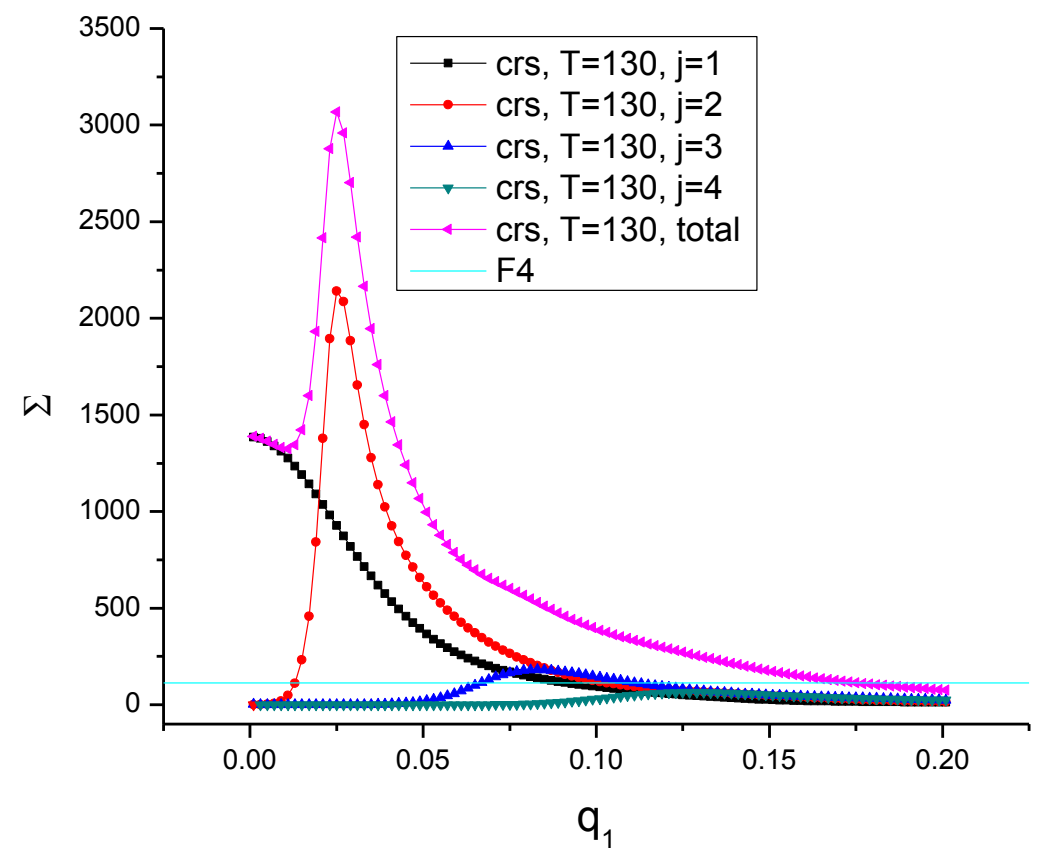

Fig.6a. QP scattering by PNR.R=6, T=130 K.

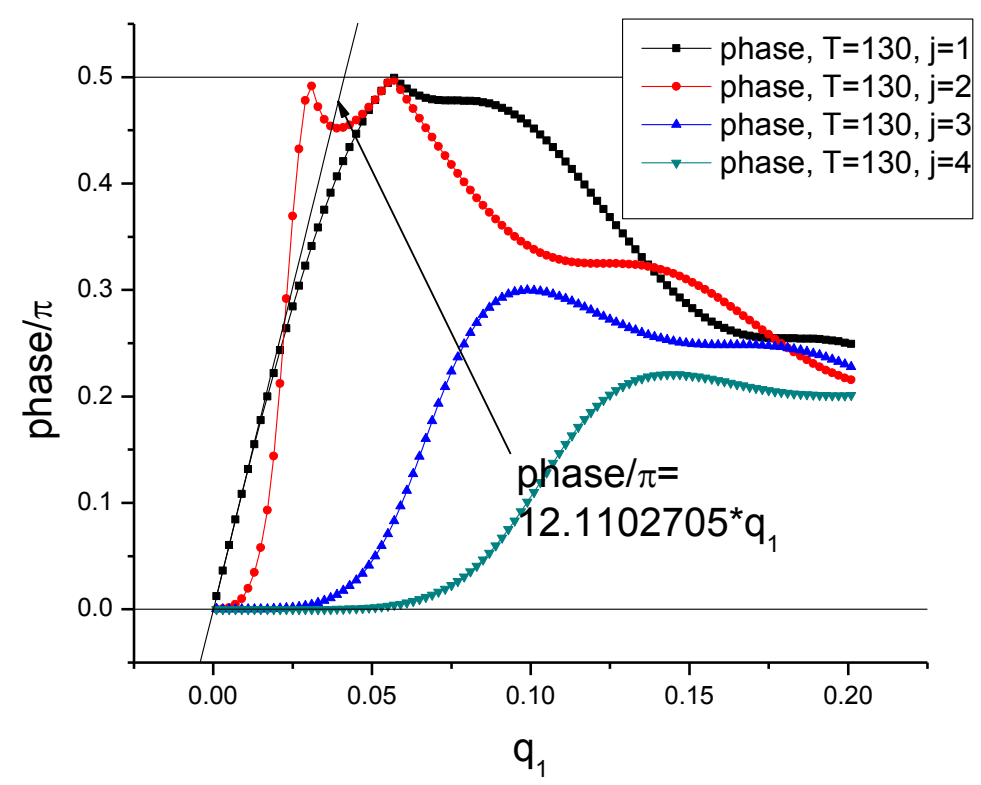

Fig.6b. Phase of QP scattering by PNR. R=6, T=130 K.

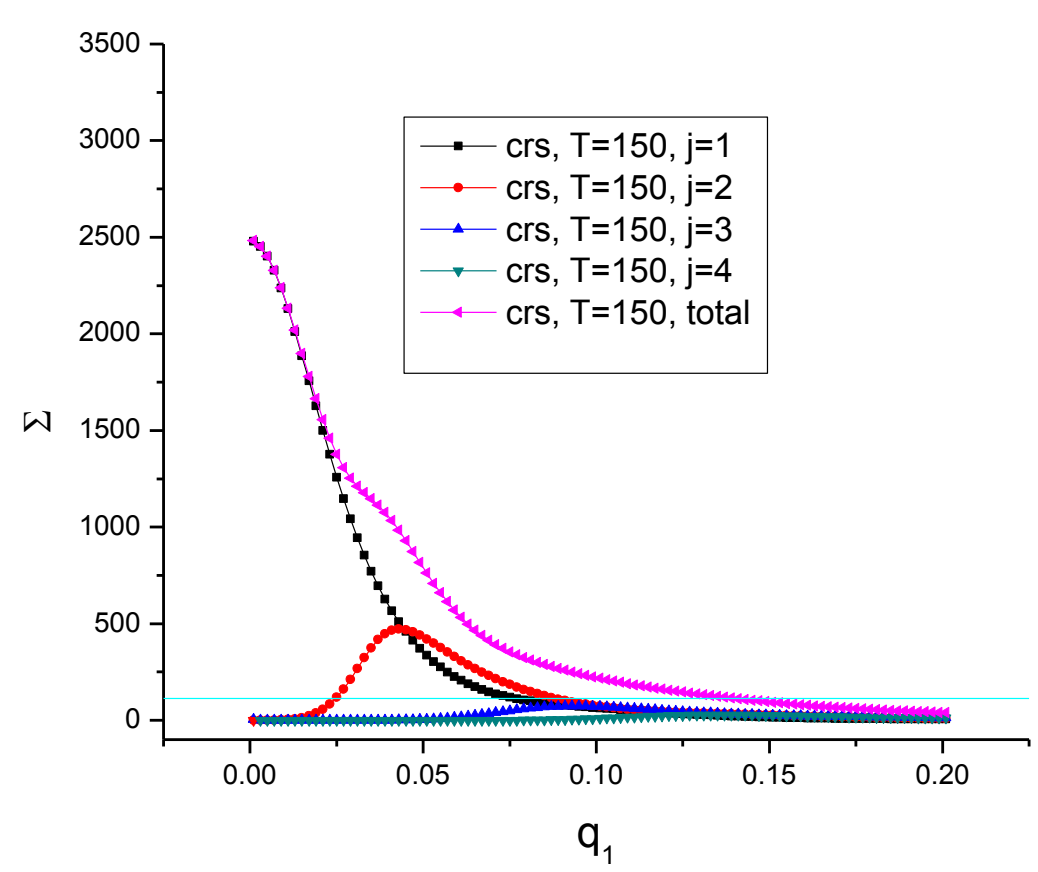

Fig.7a. QP scattering by PNR.R=6, T=150 K.

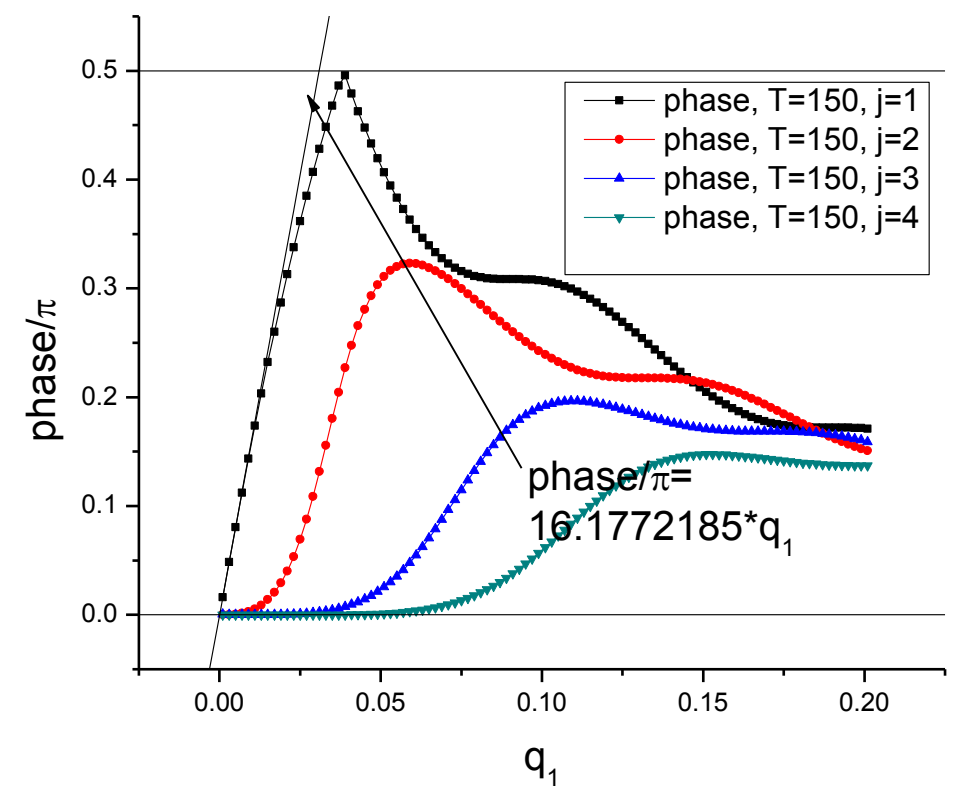

Fig.7b Phase of QP scattering by PNR. R=6, T=150 K.

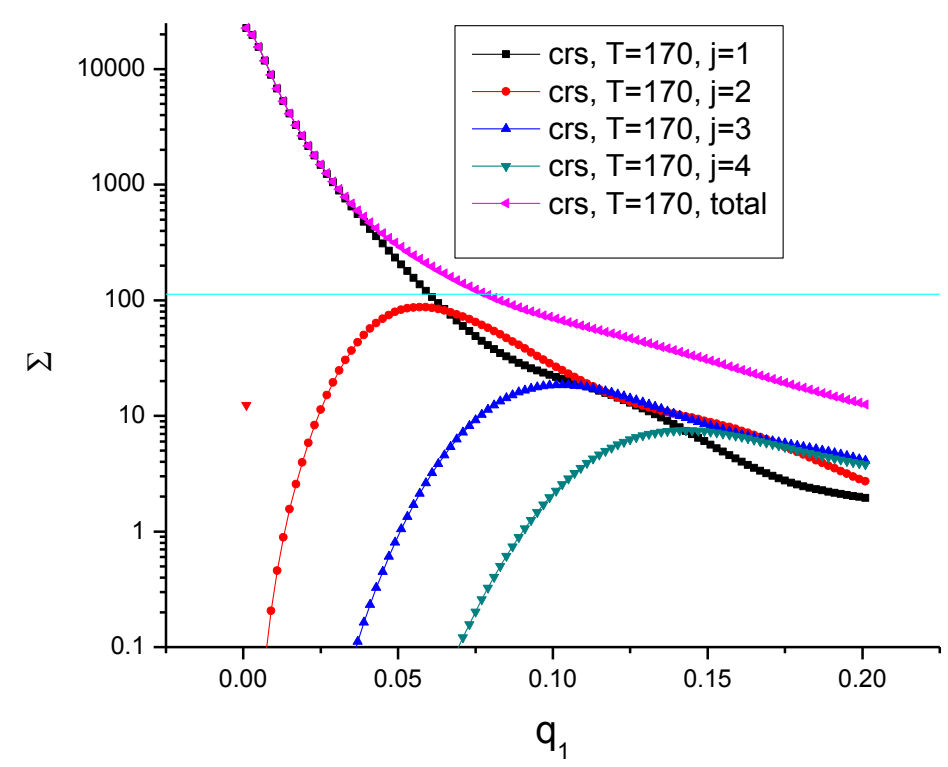

Fig. 8a. QP scattering by PNR. R=6, T=170 K, log scale.

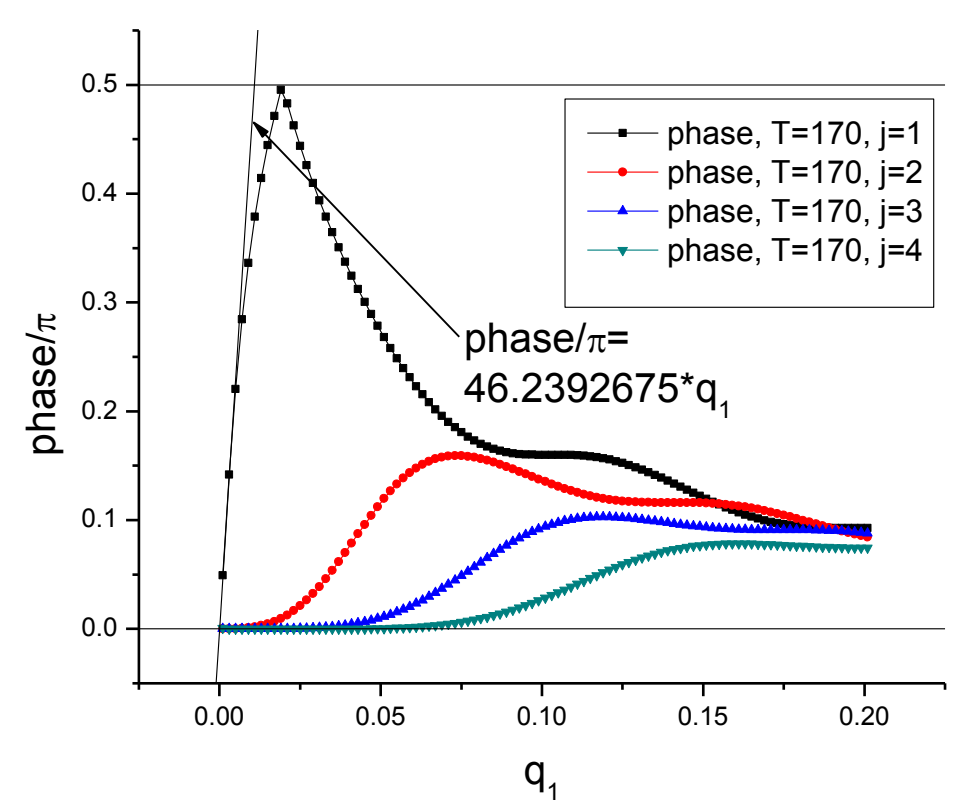

Fig.8b Phase of QP scattering by PNR. R=6, T=170 K.

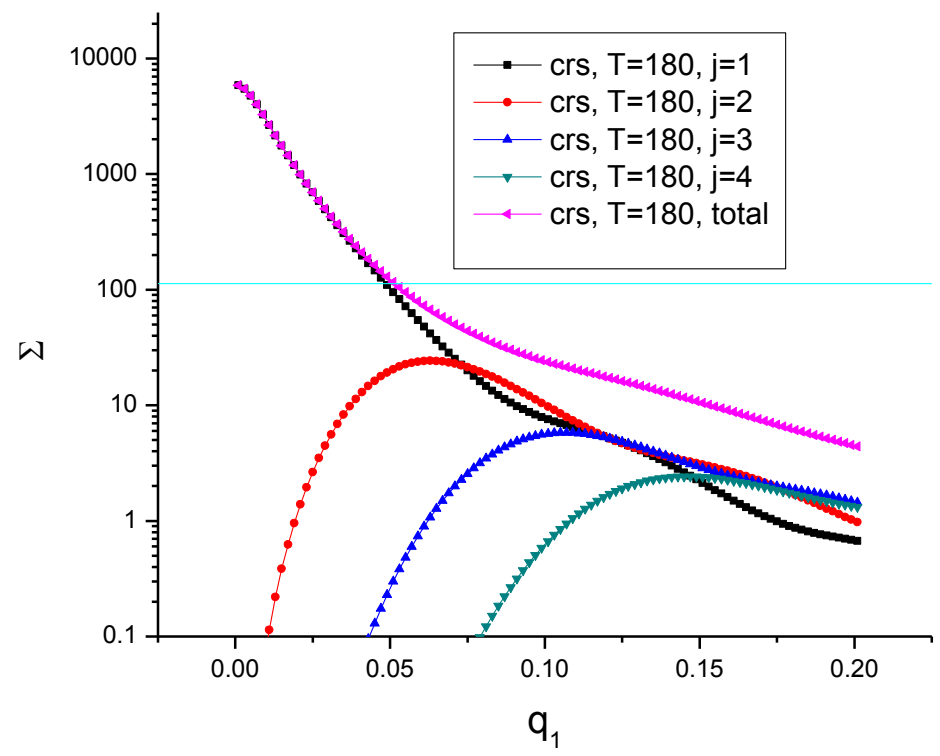

Fig.9a. QP scattering by PNR (total and partial section Σ). R=6, T=180 K, log scale.

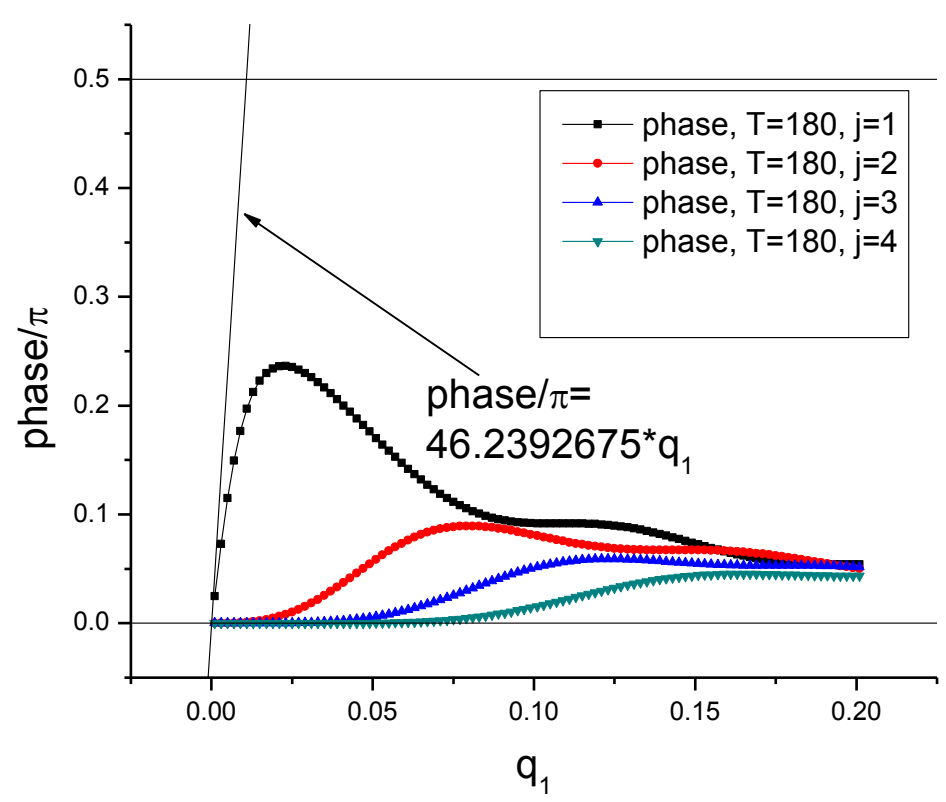

Fig. 9b Phase of QP scattering by PNR. R=6, T=180 K.

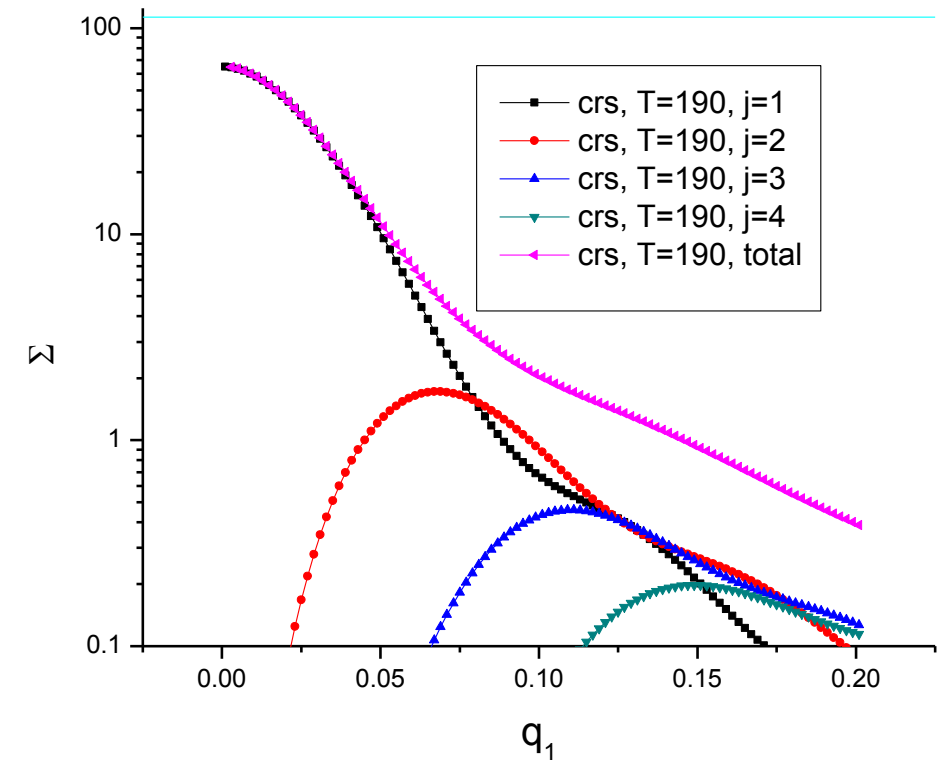

Fig.10a. QP scattering by PNR.R=6, T=190 K, log scale.

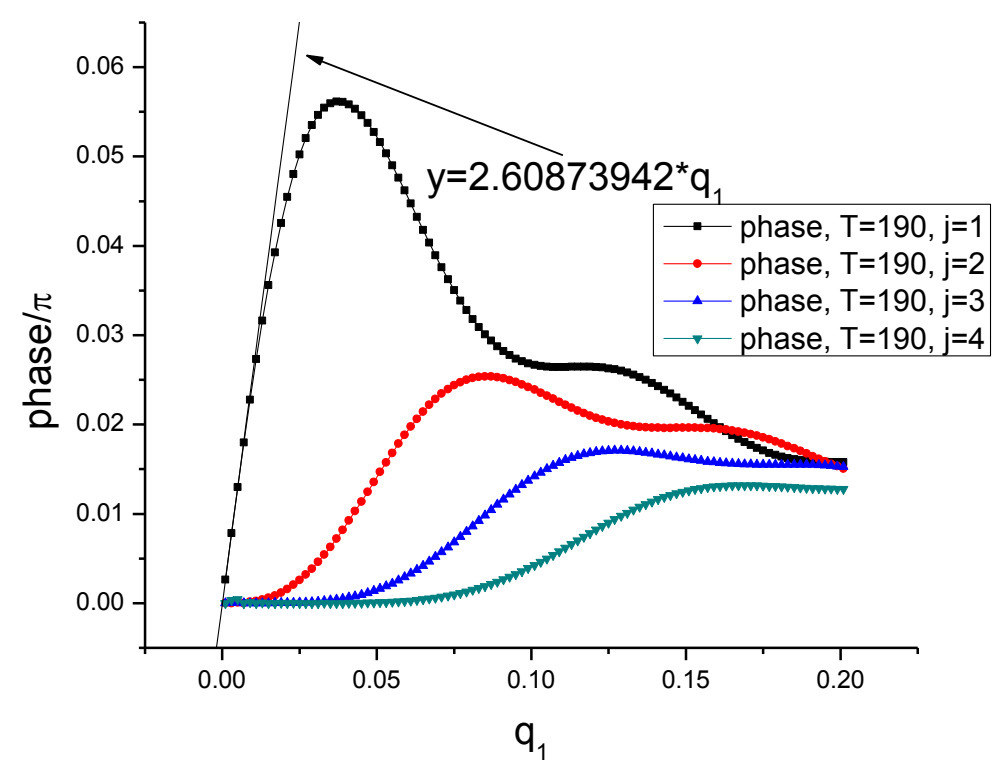

Fig.10b Phase of QP scattering by PNR. R=6, T=190 K.

## 4. DISCUSSION

We took into account depolarization field effects in our model mean field calculations of soft mode dynamics in relaxors. Depolarization field leads to the strong decreasing of the value of the temperature of the local phase transition, $\boldsymbol{T_{inst}}$, for the case of quasi polar, QP, excitations. It's happed because DF is directed against the direction of the electric dipoles and therefore leads to the increasing of the free energy value. This result is in a qualitative agreement with the molecular dynamic calculations [20 – 22]. DF effect is absent for the case of Vortex, V, excitations. Suppose that some PNR with radius ***R*** appears in process of sample cooling. If DF effect is suppressed due to the presence of the impurity charges and so on reasons then QP condensation will appears in the process of further cooling [15]. Vortex condensation could exist at the lower temperature, $VT_{instt}$, $<QPT_{inst}$ . Probably it is not contradicted to the most experimental data – PNR is described as polarized pieces of crystal. This sequence of events will be change for the case of non-screened DF: Vortex condensation will appears in the process of cooling (Fig.2) if PNR radius ***R***>8 l.u. QP condensation could exist at the lower temperature, $VT_{instt}$, > $QPT_{inst}$.

The calculated results concerning soft-mode scattering by PNR are shown in Figs. 5 - 10 and can be qualitatively interpreted similar to [14, 15] (with the important and later discussed exception - QP scattering at ***j***=1). For the case of very small incident momentum $q_1$, the TOW dynamics is similar to the scattering of a slow particle in quantum mechanics [17]. The calculated values of the QP (at j≥2) and Vortex scattering cross-section at j≥1 are in agreement with the general theory of scattering [17].

The scattering cross section Σ decreases at large momentum, $q_1$>0.15.

The value of the scattering cross section has a wide maximum in the intermediate region around $q_1 \sim 0.05 \div 0.1$. The reason for the decrease of the cross section at large momentum is as follows. Scattering is due to the difference in the soft mode gaps (eq.5) inside and outside the PNR, $\Delta\omega \equiv \omega_0(1) - \omega_0(2)$, which is increased at low temperature. This potential scattering is stronger at low temperature and decreasing for large values of $q_1$. The effect of scattering disappears at the temperature $T_{transp}$. (see Fig1) at which the soft-mode gaps inside and outside the PNR become the same and the soft-mode is no longer scattered by the PNR.

The difference between QP and Vortex scattering is essential in the resonance region due to the effective long time multiply interaction between excitation and PNR. QP and V cross sections are similar outside of this region where the corresponding cross section can be calculated in the Born approximation and with a plane wave basis. QP scattering resonances are explained (similar to [14, 15]) by the presence of quasi localized QP states.

Let’s consider more detailed QP ***j***=1 scattering by PNR. It could be noted that value of the corresponding cross section $\sigma(j=1)$ is large and saturated at the small value of $q_1$, $q_1 \approx 0.001$ *(Fig. 5-10)*. Is it consistent with law of slow particle scattering $\sigma(j=1) \sim q_1^4$ in the limit $q_1 \to 0$ ? We calculated cross section, $\sigma_{Taylor}$, at the ultra small value of $q_1$ by the expansion over $q_1$ coefficients of the eq-ns described boundary conditions and following decision of these eq-ns. Taylor (series) expansion up to 10$^{th}$ order was fulfilled by means of software Maple 10. We found that this procedure satisfactory approximates exact numerical calculation up to $q_1$=0.01 ÷0.1 (Fig.11a) and applied this approach for calculation at the extremely small value of $q_1$.

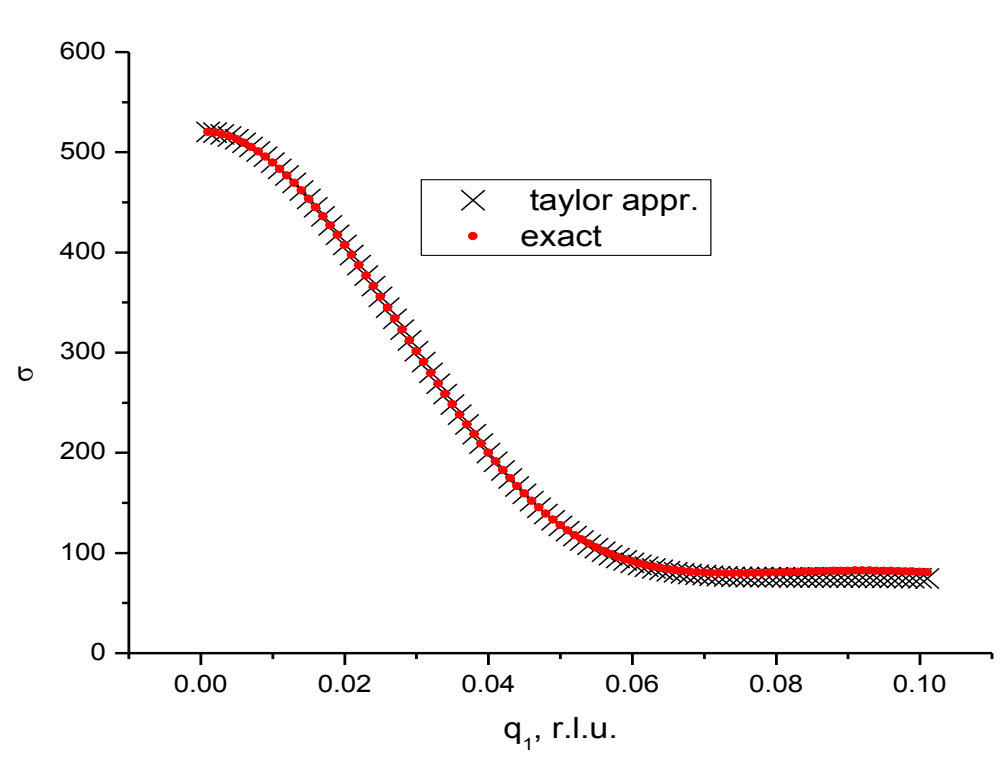

Fig.11a. Comparison of the values of the calculated QP scattering cross-section (***j***=1, ***T***=100 K, ***R***=6): exact numerical and low ***q1*** taylor (series) expansion (up to $q_1^{10}$ order) of the functions including to the boundary conditions.

Whole picture of scattering at the very small value of $q_1$ is shown at the Fig. 11b, 11c.

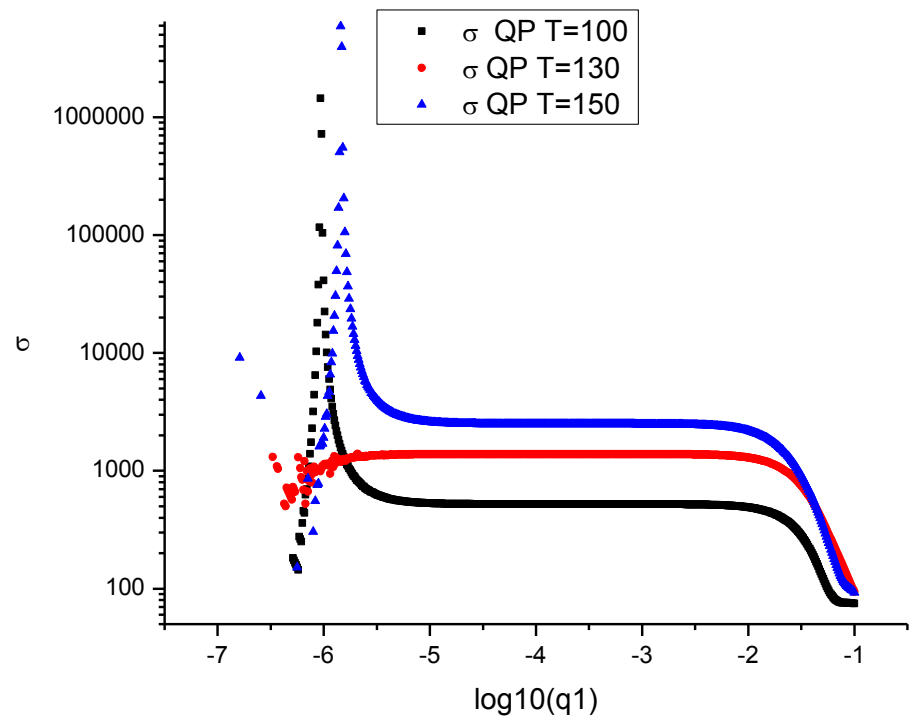

Fig. 11b.

Fig.11b. Temperature dependence of the whole picture of the low $q_1$ QP scattering, ***T***=100, 130, and 150 K; ***j***=1, ***R***=6, $\boldsymbol{\Omega_p}$=20 meV. Scattering cross section σ is shown in the log scale.

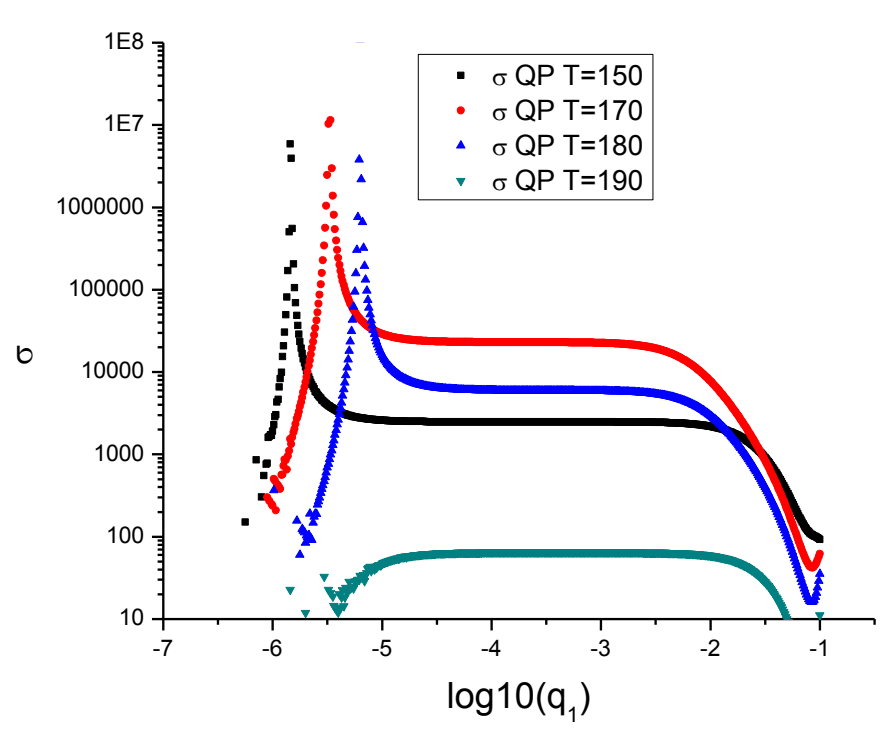

Fig.11c Temperature dependence of the whole picture of the low $q_1$ QP scattering, ***T***=150, 170, 180, and 190 K; ***j***=1, ***R***=6, $\boldsymbol{\Omega_p}$=20 meV.

Plateau cross section is small at T=190K, that is near $T_{transp}$ ≈ 194.18K, approached maximum at T≈170K and decreasing at T≈100K. The reason of that maximum existence is following. The resonance QP scattering (***j***=1) appears at the small $\boldsymbol{q_1}$ that leads to the stronger scattering at T≈170K (see Table and Fig-s 5b-10b).

Table

| T,K | Max phase/π | $q_1$ value corresp. phase maximum |
|---|---|---|
| 96 | 0.5 | 0.125 |
| 130 | 0.5 | 0.057 |
| 150 | 0.5 | 0.039 |
| 170 | 0.5 | 0.019 |
| 180 | 0.0236 | 0.023 |
| 190 | 0.056 | 0.037 |

We notice the presence of ultra narrow extremely strong peaks at the Fig 11b, 11c. These peaks describe ultra long TO wave scattering by PNR. An example of similar peak is shown in more details at the Fig. 12.

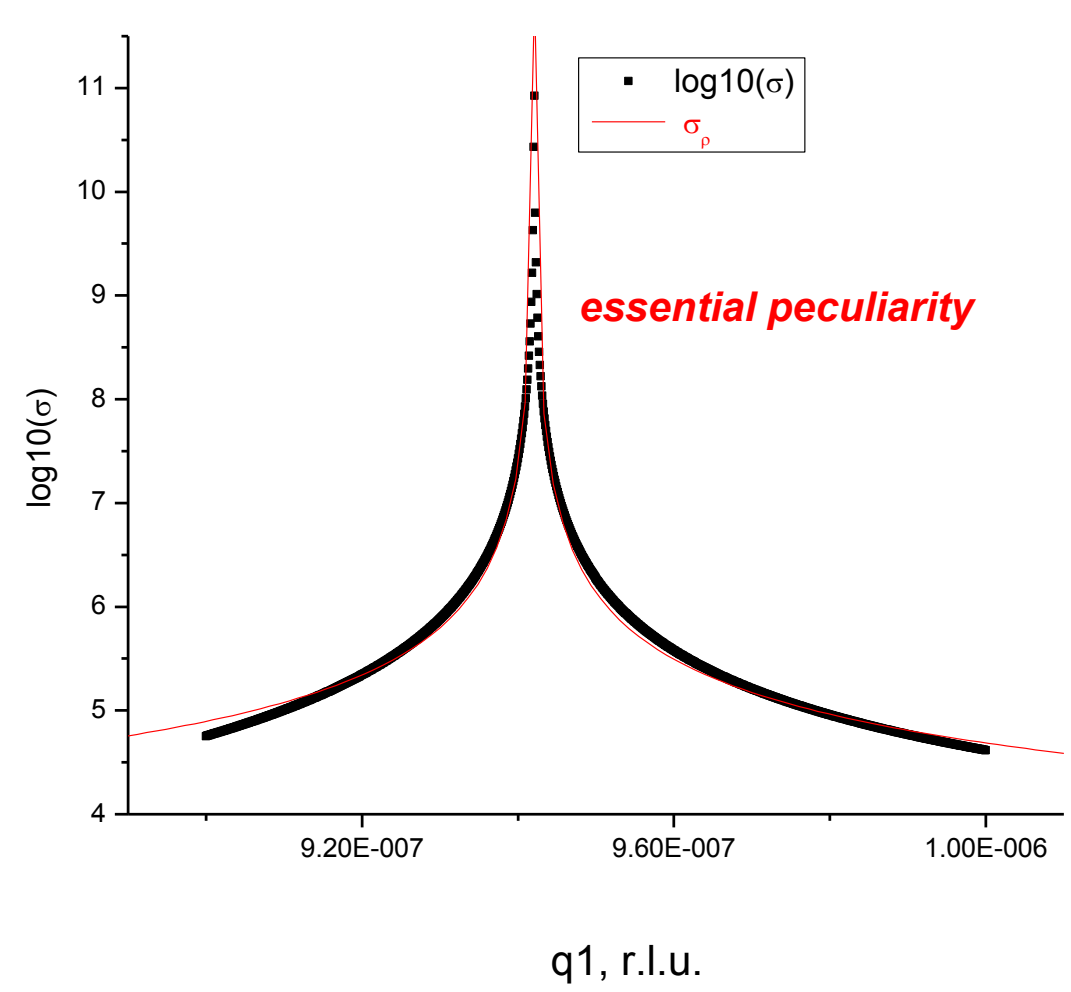

Fig.12. Ultra narrow extremely strong peak described very long QP TO wave scattering by PNR: j=1, R=6, T=100 K.

This peak shape is approximated by the expression (34) contained essential peculiarity,

$$\sigma_r \approx \exp\left\{\frac{1.128}{|q_1 - q_{1c}|^{0.135}}\right\}, \quad q_{1c} = 9.42e-07 \ (34)$$

Peak has very wide winds, it shape isn't similar to the well known Lorentzian or Gaussian. Extremely small value of the peak position center $\boldsymbol{q_{1c}}$ is corresponding to the TOW wave length equal to the value of several mm. Cross section of scattering is decreased below of this resonance in an agreement with general theory [17].

Ultra narrow resonance peak center displaces to the large value of $\boldsymbol{q_1}$ with increasing temperature. This peak is diminished at T=190K. Probably this extremely narrow resonance peak is difficult to observe, it could be destroyed by sample imperfections. However here we are interesting mainly by effect of the very far wings, corresponding to the value of the cross section to the several orders smaller than one in the resonance center. Therefore we could expect that sample imperfections effect will be no strong in this case.

TOW plateau scattering parameters are dependent of course vs. PNR size (Fig.13).

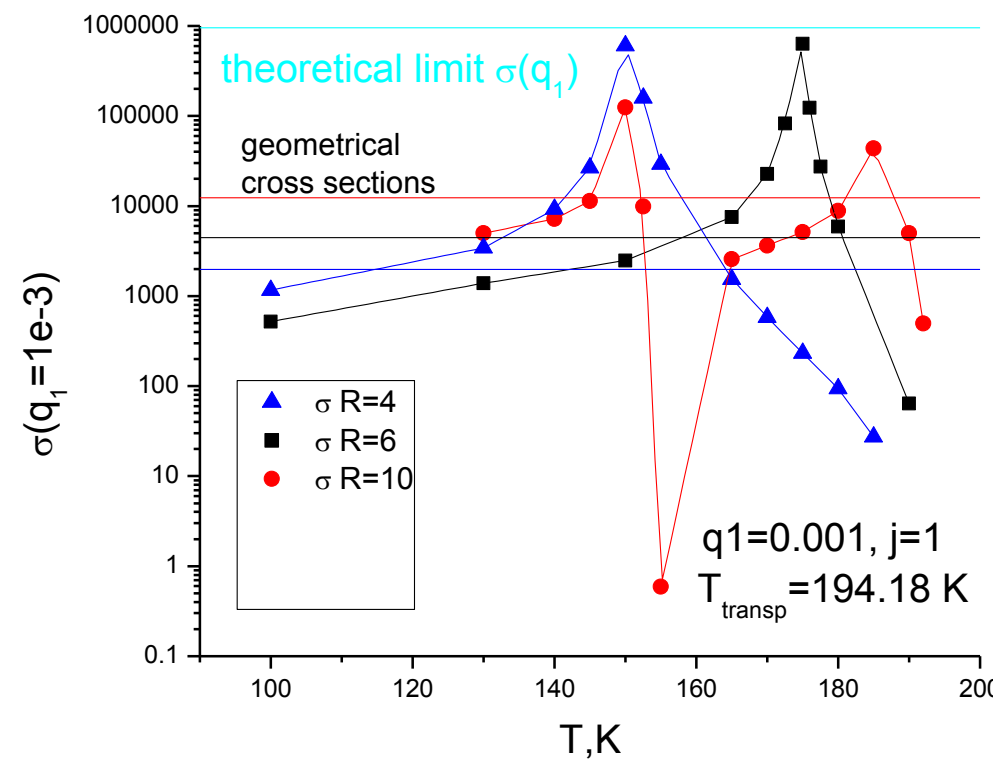

Fig.13. Temperature dependence of TOW plateau scattering by PNR: $\boldsymbol{q_1}$=0.001, $\boldsymbol{R}$=4, 6, and 10; $\boldsymbol{j}$=1, $\boldsymbol{\Omega_p}$=20 meV.

Effective TOW attraction to PNR is more for the case of large size PNR. Therefore cross section peak at $\boldsymbol{R}$=6 is closer to the $\boldsymbol{T_{transp}}$ than peak corresponding $R$=4. Two peaks corresponding R=10 appearing due to the existence of different mode of motion.

The existence of the strong TOW scattering (plateau) at $\boldsymbol{q_1} \leq 0$ .01 could be explained as following. The components of the displacement outside of PNR ***wfp1, wfr1,*** perpendicular and parallel to PNR radius respectively could be written in the form:

$$wfp1 = \frac{4\pi i A in e^{(-iq_1 R)}((q_1 R)^2 - q_1 R - 1)}{(q_1 R)^3} - \frac{4\pi i A out e^{(+iq_1 R)}((q_1 R)^2 + q_1 R - 1)}{(q_1 R)^3} + \frac{C_{1D}\sqrt{2}}{R^3},$$

$$wfr1 = \frac{4\pi i\sqrt{2} A in e^{(-iq_1 R)}(q_1 R + 1)}{(q_1 R)^3} + \frac{4\pi i\sqrt{2} A out e^{(+iq_1 R)}(q_1 R - 1)}{(q_1 R)^3} - \frac{C_{1D} 2}{R^3} \quad (35a)$$

Here ***Ain***, ***Aout***, $\boldsymbol{C_{1D}}$ – amplitude of in, out wave and depolarization displacement respectively. We omitted here contribution of the LOW displacement concentrated at the narrow layer at the PNR surface and (for simplicity) the angular part of the expressions.

We have by after serial expansion over $\boldsymbol{q_1}$:

$$wfp1ser = 2\pi i\left[\frac{2}{(q_1 R)^3} - \frac{1}{q_1 R}\right](-Ain + Aout) + \frac{8\pi}{3}(Ain + Aout) + \frac{C_{1D}\sqrt{2}}{R^3},$$

$$wfr1ser = -2\pi i\left[\frac{2}{(q_1 R)^3} + \frac{1}{q_1 R}\right]\sqrt{2}(-Ain + Aout) + \frac{4\pi}{3}\sqrt{2}(Ain + Aout) - \frac{C_{1D} 2}{R^3} \quad (35b)$$

Let's suppose for a moment that we neglect depolarization displacement, that is $\boldsymbol{C_{1D}}$=0. In order to have the finite value of the displacement at the PNR surface, we should propose that $\boldsymbol{A_{out} \approx A_{in}(1+\sim q_1^3)}$, so that the phase shift of the reflected wave will be ~ $q_1^3$, and cross section of scattering ~ $\boldsymbol{q_1}^{(6-2)}$~ $\boldsymbol{q_1}^4$ in agreement with general theory of scattering [17]. However $\boldsymbol{C_{1D}}\neq 0$ and displacements ***wfp1, wfr1*** will be finite if $\boldsymbol{C_{1D} \sim 1/ q_1^2}$, $\boldsymbol{A_{out} \approx A_{in}(1+\sim q_1^1)}$ (this is confirmed by the numerical calculation at the $q_1 \ll 1$), phase shift ~ $\boldsymbol{q_1}$ and cross section of scattering is independent vs $q_1$ (plateau). Dependence $\boldsymbol{C_{1D} \sim 1/ q_1^2}$ is in agreement with expression (24b) and demonstrates large susceptibility of the DM displacements for the respect of the electric depolarization field at the small value of $\boldsymbol{q_1}$.

Appearance of the scattering cross section plateau at ***j***=1 leads to the serious sequences. Let's suppose that C is the PNR concentration. TOW mean free pass length $\lambda_{mfp} = \frac{1}{C\sigma(q1)}$ ***;*** mean free pass time $\boldsymbol{\tau_{mfp}} = \frac{\lambda_{mfp}}{V_{gr}}$, group velocity $V_{gr} \approx \frac{C_0^2}{\omega_0(1)} q_1$ and $\frac{\boldsymbol{\tau_{mfp}}\omega}{2\pi} = \frac{1}{2\pi q_1} \frac{\omega_0^2(1)}{C_0^2 C\sigma}$. We should require that $\lambda_{mfp} > 2\pi/q_1$ in the frame of ordinary picture of spreading TOW. However this picture of the spreading TOW will be violated at the small value of $\boldsymbol{q_1 = q_0}$ due to the presence of plateau at the curve $\sigma(\boldsymbol{q_1})$ if PNR concentration $C > \frac{q_1}{2\pi\sigma}$ . For an example, if $q_0$=0.01, σ=1000, C>1.6e-06. The limit of non-overlapping PNR concentration $C_{lim}=1/(4\pi/3*(R)^3)$=1.1 e-03. Therefore we will be near the region of non-spreading soft mode in this case. We have TOW over damping regime for the case $\frac{\boldsymbol{\tau_{mfp}}\omega}{2\pi} < 1$. This over damping condition is different from non-spreading TOW condition. Therefore we can have non-over damping localized or quasi localized soft mode. Of course the strong soft-mode scattering leads to the some effect on the TOW dispersion too. We hope to consider the complicated effect of dispersion in more details separately.

Gehring et al [9] attributed waterfall results to a sharp step-like increase in the TO damping for the momentum $q<q_0$ . However here could be some trouble.

TOW is corresponded to the angular momentum $\boldsymbol{j \geq 1}$ for the case of isotropic medium. Therefore cross section, $\boldsymbol{\sigma}$, of the TO scattering by PNR should be small at the small momentum ***q*** and waterfall should be suppressed in this case. Calculation results [14, 15] demonstrate that TO scattering by PNR is really strongly increased at small ***q*** but nullified at, say, ***q***<0.01 r.l.u. in agreement with general theory [17]. Depolarization field effect also leads to the result $\boldsymbol{\sigma \sim q^4}$ at ***j=1*** but only at the extremely small momentum, ***q<5E-7*** r.l.u., corresponding to the size 1/q ~100 µm. It is happened due to the following. Depolarization field interact with optical mode and created corresponding displacements, which could be titled as Depolarization mode (DM). DM space dependence ($\sim \boldsymbol{r}^{j-1}$ inside PNR and $\sim \boldsymbol{r}^{-j(j+1)-1}$ outside of PNR) is quite different from ~sin(***qr-ωt***) corresponding TO wave. The incident TO wave excites charge fluctuations at the PNR surface. These charge fluctuations create DF interacted with optical mode displacements and leading to the increasing of scattering. Scattering cross section $\boldsymbol{\sigma}$ corresponding angular harmonic ***j***=1 is steadily increased parallel to the decreasing value of momentum ***q*** up to the very large resonance value $\boldsymbol{\sigma}$ at $\boldsymbol{q=q_{res}}$ and crossing to the dependence $\boldsymbol{\sigma \sim q^4}$ ***at*** $\boldsymbol{q< q_{res}}$ . This resonance is extremely narrow. For example value of $\boldsymbol{q_{res}}$=5E-7 r.l.u. and resonance energy width ~1E-12 meV~1Hz. Probably observation of similar resonances is out of limit today experimental technique, taking into account problems with resolution, sample quality and so on. Scattering at ***q***~0.01÷0.2 r.l.u. is very far from resonance and so no sensitive for the quality of sample.

Therefore plateau of scattering is existed and above mentioned trouble is over.

Real waterfall was observed [9] at the momentum $\boldsymbol{q_1} \leq q_0$, $q_0$~ 0.1. The value of $q_0$ is much more than beginning of plateau of scattering calculated above. Therefore model parameters optimization and crystal anisotropy should be taken into account for the direct comparison with the existing experimental data.

It's almost impossible to observe electric DF by means of thermal neutron scattering. However depolarization mode displacements are excited due to the scattering of the "ordinary" thermal TO excitations at the PNR boundaries and probably could be observed by means of inelastic nuclear neutron scattering near the Bragg peaks in crystals.

ACKNOWLEDGMENTS

I would like to express my deep gratitude to the Prof. J.Toulouse who introduced me to the modern relaxor research.